\documentclass[draft]{agujournal2019}
\usepackage{url} 
\usepackage{lineno}
\usepackage{soul}

\draftfalse
\journalname{JGR: Space Physics}

\begin{document}

\title{The Spatiotemporal Structure of Induced Magnetic Fields in Callisto's Plasma Environment due to their Propagation with MHD Modes}

\authors{David Strack\affil{1}, Joachim Saur\affil{1}}
\affiliation{1}{Institute of Geophysics and Meteorology, University of Cologne, Cologne, Germany}
\correspondingauthor{David Strack}{d.strack@uni-koeln.de}

\begin{keypoints}
      \item Low-frequency induced magnetic fields are transported by the magnetohydrodynamic modes in the plasma environments of the Galilean moons
      \item Transport effects alter the amplitude of the induced magnetic fields and cause an additional temporal delay
      \item Transport effects are consistent with the magnetic field observations of Galileo's C03 and C09 Callisto flyby 
\end{keypoints}

\begin{abstract}
	We investigate how the spatiotemporal structure of induced magnetic fields outside of Callisto is affected by their propagation with the magnetohydrodynamic (MHD) modes.
	At moons that are surrounded by dense magnetized plasmas like the Galilean moons, low-frequency induced magnetic fields cannot propagate with the ordinary electromagnetic mode as is implicitly used by standard analytical expressions.
	Instead, the induced magnetic fields propagate with the MHD modes, which exhibit anisotropic propagation properties and have finite velocities.
	Using an MHD framework, we model the spatiotemporal effects of the transport on the induced signals and analyze their contribution to Galileo's C03 and C09 flyby observations.
	We find that the induced magnetic field in Callisto's plasma environment is asymmetric with a pronounced upstream/downstream asymmetry.
	By neglecting the transport effects, the amplitude of the induced magnetic field is under- or overestimated by up to tens of percent, respectively.
	Additionally, we find that MHD wave and convection velocities are strongly reduced in Callisto's local plasma environment, resulting in an additional temporal delay between the emergence of the induced field at the surface of Callisto or the top of its ionosphere and the measurements at spacecraft location.
	The associated phase shift depends on the location of the observer and can reach values of several to tens of degrees of the phase of the primary inducing frequency.
	Transport effects impact the observed induction signals and are consistent with the C03 and C09 magnetic field measurements.
\end{abstract}

\section*{Plain Language Summary}

Jupiter's moon Callisto experiences a time-variable magnetic field that interacts with Callisto and causes a so-called induced magnetic field.
The strength, shape, and change over time of the induced magnetic field depend on the electrically conductive layers of Callisto, such as its potential subsurface ocean or the conductive part of its atmosphere.
Therefore, these conductive layers can be studied by analyzing the magnetic fields measured by the Galileo spacecraft during close encounters with Callisto.
In this study, we use computer simulations to investigate how the induced magnetic field is affected by its transport from the surface of Callisto to the spacecraft hundreds of kilometers away.
We find that the transport in Callisto's space environment has a considerable impact on the induced magnetic field.
Depending on the side of Callisto, the strength of the induced magnetic field is reduced or increased by tens of percent.
Additionally, the transport causes a time delay between the generation and measurement of the induced magnetic field.
With this study, we contribute an important aspect to the analysis of the induced magnetic field, relevant for assessing whether Callisto has a subsurface ocean and relevant for the future Europa Clipper and JUICE satellite missions to Jupiter's moons.


\section{Introduction}
\label{sec:introduction}

The first three flybys of the Galileo spacecraft at Callisto revealed induced magnetic fields \cite{Khurana1998,Neubauer1998a,Kivelson1999,Zimmer2000,Liuzzo2016}.
Yet, although these flybys, commonly referred to as C03, C09 and C10, occurred more than two decades ago, the origin of these induced magnetic fields is still not fully understood.
While early studies favored the existence of a water ocean underneath Callisto's icy crust \cite{Khurana1998,Neubauer1998,Kivelson1999,Zimmer2000}, \citeA{Hartkorn2017b} argued that Callisto's ionosphere plays an important role and might be sufficiently conductive to be the primary origin of the observed induction signals.
One of the challenges in interpreting the induced magnetic fields is that Callisto is embedded in Jupiter's magnetospheric plasma, which strongly affects the induced and ambient magnetic fields.
With this study we further contribute to the understanding of induced magnetic fields at Callisto and in its plasma environment.

In its rest frame, Callisto experiences a quasi-periodically time-varying magnetic field with a period of ${\sim}10\,$hours, the primary cause being Jupiter's inclined dipole.
The time-varying magnetic field induces electric fields that cause electric currents in conductive layers.
These currents, in turn, are the source of secondary (induced) magnetic fields that may be used to analyze the conductive layers. 
However, while such secondary magnetic fields have been observed at all of the icy Galilean moons \cite{Khurana1998,Kivelson1999,Kivelson2000,Kivelson2002,Zimmer2000,Saur2015}, the interpretation of induced magnetic fields can be challenging. 
First, with only a few spacecraft flybys, the available data is sparse.
Second, interactions between the flow of the magnetospheric plasma and the moons additionally perturb the magnetic field.
Depending on Callisto's position with respect to Jupiter's magnetic equatorial plane and to the plasma sheet, these perturbations can be of equal or even larger magnitude than the induction signals \cite{Liuzzo2015}.
Although different in origin, these moon-magnetosphere interactions can not only obscure but sometimes even resemble the signatures of an induced magnetic field.
Due to their non-linear and coupled nature, moon-magnetosphere interactions are most accurately described by numerical modelling. For Callisto, such modelling was carried out by \citeA{Seufert2012}, \citeA{Lindkvist2015}, and \citeA{Liuzzo2015,Liuzzo2016,Liuzzo2017}.

In case an induced magnetic field can be identified in the measured data, it is the result of three contributions: the properties of the inducing (primary) magnetic field, the properties of the moon's conductive structures, and the induced magnetic field's propagation from the conductor to the observing spacecraft. 
Each of these contributions involves uncertainties and influences the measured magnetic field.
Thus, for the interpretation of the data, a qualitative as well as quantitative understanding of these contributions is required.
Yet, previous studies focused on the inducing magnetic field and/or a moon's conductive structures \cite{Zimmer2000,Schilling2007,Seufert2011,Vance2021,Styczinski2022,Winkenstern2023} and neglected the induced magnetic field's propagation.
With this study, we aim to fill this void.
Numerical moon-magnetosphere interaction models that account for induced magnetic fields include the transport and modification of the induced magnetic field in a moon's plasma environment if the magnetic boundary condition at the surface of the moon is correctly chosen.  
This is reflected, for example, in the numerical results of \citeA{Liuzzo2016}, who disentangle plasma interactions and induction signatures for Galileo's C10 flyby.
In this study, however, we explicitly investigate and quantify the transport effects on the induced magnetic field. 

Regardless of whether the induced magnetic fields originate from Callisto's interior or its ionosphere, Galileo observed these fields at a distance from their origin.
Thus, the induced magnetic fields needed to propagate from the conductor to the magnetometer.
As we will argue in Section~\ref{sec:motivation}, the induced magnetic field cannot propagate with the ordinary electromagnetic wave mode in Callisto's plasma environment but instead, due to its low frequency, it propagates with the magnetohydrodynamic (MHD) modes.
Because of the anisotropic propagation properties of the MHD modes, the propagation has a spatiotemporal effect on the observable induced magnetic field, which has not yet been explicitly analyzed.
Therefore, we investigate the spatiotemporal modifications to the induced magnetic field as it is transported by MHD modes in Callisto's space plasma environment.
For this investigation, we use an MHD model (introduced in Section~\ref{sec:model}) to simulate the propagation and non-linear coupling of the MHD waves.
We apply this model to two scenarios: First, to a scenario with simplified geometry to analyze the overall spatiotemporal implications (Section~\ref{sec:results}), and second, to the C03 and C09 flyby scenarios as a case study (Section~\ref{sec:galileo}).
In Section~\ref{sec:discussion}, we discuss how additional sources of asymmetry could affect our results and consider the relevance of transport effects at the other Galilean moons.
Finally, Section~\ref{sec:conclusion} summarizes the main results.


\section{Modes of Propagation of Induced Magnetic Fields}
\label{sec:motivation}

Since the spatiotemporal properties of induced magnetic fields depend on the properties of the conductive layer, such as its conductivity or spatial extent, magnetic field measurements by space probes can be used to constrain these properties.
Assuming a conductivity model for the planetary body, the magnetic field that would hypothetically arise from this model is calculated.
The misfit between calculated and measured data then allows to judge the plausibility of the selected conductivity model.  
Inaccuracies in this calculation may, in turn, lead to misinterpretations about the planetary body.

For simplified scenarios, several analytic solutions exist in the literature for the forward calculation of induced magnetic fields outside the planetary body.
For the details, we refer the reader to the works, e.g., of \citeA{Srivastava1966}, \citeA{Parkinson1983}, \citeA{Zimmer2000}, \citeA{Saur2010}, \citeA{Seufert2011} and \citeA{Grayver2024}.
These solutions describe the planetary body to consist of one or several spherical symmetric shells of constant conductivity surrounded by a resistive, non-conducting medium such as the icy shell of Callisto or free space.
Under these conditions, it is possible to express the induced (secondary) magnetic field $\mathbf{B}_\mathrm{sec}$ in relation to the dipole field $\mathbf{B}_\mathrm{sec}^\infty$ that would be induced by a spatially homogeneous inducing (primary) magnetic field $\mathbf{B}_\mathrm{prim}$ varying at frequency $\omega_\mathrm{prim}$ within a perfectly conducting sphere of radius $R_\mathrm{C}$:
\begin{linenomath*}
\begin{equation}
	\label{eq:motivation_Bsec}
    \mathbf{B}_\mathrm{sec} \left(\mathbf{r}, t\right) 
    	= A \cdot \mathbf{B}_\mathrm{sec}^\infty \left(\mathbf{r}, t - \frac{\phi^\mathrm{ph}}{\omega_\mathrm{prim}}\right),
\end{equation}
\end{linenomath*}
\begin{linenomath*}
\begin{equation}
  	\label{eq:motivation_Bsec-infty}
    \textrm{with }
    \mathbf{B}_\mathrm{sec}^\infty \left(\mathbf{r}, t \right)
    	= \frac{\mu_0}{4 \pi}
			\left(
				3 \left(\mathbf{r} \cdot \mathbf{M}\left(t\right) \right)\mathbf{r} 
        		- r^2 \mathbf{M}\left(t\right) 
    		\right) / r^5
    ,\quad
    \mathbf{M}\left(t\right)
		= \frac{2\pi}{\mu_0} \mathbf{B}_\mathrm{prim}\left(t\right) R_\mathrm{C}^3 .
\end{equation}
\end{linenomath*}
In this relation, $A$ is a factor that controls the amplitude of the induced magnetic field and the phase shift $\phi^\mathrm{ph}$ quantifies the time-delay between the inducing and induced magnetic field for the finite conductivity cases.
By using a negative sign with the phase shift in Equation (\ref{eq:motivation_Bsec}), we consider $\phi^\mathrm{ph} \geq 0$ (see also \citeA{Vance2021}).
Throughout this study, $t$ and $\mathbf{r}$ denote time and position, with $r = |\mathbf{r}|$.

In these analytical solutions, it is assumed that the resistive layers and the space environment of the planetary bodies are free of current.
This assumption implies that within these regions, the induced magnetic field propagates with the isotropic electromagnetic wave mode of free space.
By also disregarding the displacement currents in Ampere's law, this results in the induced magnetic field being instantaneous everywhere (i.e., with a hypothetically infinite wave velocity).
However, Callisto and the other Galilean moons are embedded in Jupiter's magnetospheric plasma.
In this plasma, the isotropic ordinary electromagnetic wave mode cannot propagate at the low wave frequencies $\omega$ corresponding to periods of 10 hours.
Instead, the induced magnetic field is propagated by low frequency MHD wave modes.

In a plasma, a variety of different wave modes exists.
However, depending on the properties of the plasma, only a subset of these wave modes propagates at a given frequency $\omega$ \cite<see e.g.>[]{Stix1992}.
The subset of propagating wave modes in a cold plasma can be roughly classified based on characteristic plasma frequencies: the electron and ion plasma frequencies $\omega_\mathrm{pe}, \omega_\mathrm{pi}$ as well as the electron and ion cyclotron frequencies $\omega_\mathrm{ce}, \omega_\mathrm{ci}$.
\begin{linenomath*}
\begin{equation}
    \label{eq:motivation_plasma-frequency}
    \omega_\mathrm{pe}^2 = \frac{n_e e^2}{\epsilon_0 m_e}, \quad \omega_\mathrm{pi}^2 = \frac{n_i q^2}{\epsilon_0 m_i}
\end{equation}
\end{linenomath*}
\begin{linenomath*}
\begin{equation}
    \label{eq:motivation_gyro-frequency}
    \omega_\mathrm{ce} = \frac{eB}{m_e}, \quad \omega_\mathrm{ci} = \frac{qB}{m_i}
\end{equation}
\end{linenomath*}
Here, $B$ is the magnitude of the magnetic field and $n_e, m_e, e$ and $n_i, m_i, q$ are the number density, mass, and charge of the electrons and ions, respectively.

The ordinary (O) wave mode, the equivalent to the electromagnetic wave of free space in a plasma, propagates for $\omega > \omega_\mathrm{ce}$ and $\omega^2 > \omega_\mathrm{pe}^2 + \omega_\mathrm{pi}^2$ and ceases to exist for lower frequencies. 
At the other extreme, for frequencies $\omega^2 \ll \omega^2_\mathrm{pe} + \omega^2_\mathrm{pi}$ and $\omega \ll \omega_\mathrm{ci}$, waves only propagate in the MHD-regime.

\citeA{Kivelson2004} list properties of the plasma that surrounds Callisto.
The most dilute plasma in the listed range has an electron number density of $n_e = 0.01\,$cm\textsuperscript{-3}, which corresponds to an electron plasma frequency of $f_\mathrm{pe} = \omega_\mathrm{pe}/2\pi \approx 9.0 \cdot 10^2\,$Hz.
For the ion cyclotron frequency, they list the range $1\cdot 10^{-2} - 3\cdot 10^{-1}$\,Hz.
The minimal electron plasma frequency obtained from Galileo's Plasma Wave Subsystem \cite{Ansher2017} in the vicinity of Callisto ($r \leq 10\cdot R_\mathrm{C}$, $R_\mathrm{C}$ = Callisto's radius) is $f_\mathrm{pe} = 7.2$\,Hz.  
Both values for the electron plasma frequency in Callisto's vicinity are orders of magnitude higher than the primary frequency associated with Callisto's inducing and induced magnetic field $\omega_\mathrm{prim}/2\pi = \frac{1}{T_\mathrm{prim}} < 2.8 \cdot 10^{-5}\,$Hz.
Thus, within the magnetospheric plasma that surrounds Callisto, the isotropic ordinary electromagnetic wave mode cannot propagate.
Instead, since $\omega_\mathrm{prim}^2 \ll \omega_\mathrm{pe}^2 < \omega_\mathrm{pe}^2 + \omega_\mathrm{pi}^2$ and $\omega_\mathrm{prim} \ll \omega_\mathrm{ci}$, the magnetic field perturbations that arise from the induced field must propagate with the MHD modes.
These are the compressional \textit{slow} and \textit{fast} wave modes, the transverse \textit{Alfvén} wave mode, and additionally, due to the relative motion of the magnetospheric plasma, the \textit{convection} mode.
Among these wave modes, only the fast mode propagates momentum and energy at an arbitrary angle with respect to the direction of the magnetic field.
The slow mode and the Alfvén mode do not propagate at all perpendicular to the magnetic field and transport momentum and energy primarily along the magnetic field.
The convection mode is operating only in the downstream flow direction.

Between the MHD modes and the ordinary electromagnetic mode there are two important differences in Callisto's plasma environment:
\begin{enumerate}
    \item The MHD modes exhibit strongly anisotropic propagation properties, i.e., their phase and group velocities depend on the direction of the magnetic field.
    \item The MHD modes have wave velocities significantly smaller than the speed of light.
\end{enumerate}
In this work, we systematically investigate whether the differences in the propagation properties have temporal as well as spatial implications on the induced magnetic field observable in Callisto's vicinity and how they affect the interpretation of current and future measurements to probe the ionosphere and/or interior of Callisto.

\section{Model}
\label{sec:model}

We model Callisto's space plasma environment and the propagation of the induced magnetic field using an MHD framework.
By acting as an obstacle to the flow of Jupiter's magnetospheric plasma, Callisto and its gaseous envelope perturb this flow and thereby continuously excite MHD waves.
Carried by these waves and the convection mode, the perturbations propagate away from Callisto and shape its local plasma and magnetic field environment.
In addition to moon-magnetosphere interactions, the induced magnetic field propagates as well.
Both the propagation of the induced magnetic field and the plasma perturbations result in standing waves in the rest frame of Callisto, i.e., structures with $\partial/\partial t$ of the MHD quantities being approximately zero.
To capture the impact of the induced magnetic field in this complex system of moon-magnetosphere interactions, our numerical studies are based on the following procedure:
\begin{enumerate}
    \item We perform a \textit{reference} simulation to obtain the inhomogeneous local plasma environment, shaped only by the moon-magnetosphere interactions.
    \item Based on this stationary reference model, we excite magnetic field perturbations representative of an induced magnetic field at our simulation domain's inner boundary. At the boundary layer, these perturbations correspond to the values of an induced magnetic dipole field of a given amplitude $A$.
    \item As the simulations continue, these perturbations propagate into the simulation domain. Since both the MHD wave modes and the convection mode are included self-consistently in the full set of MHD equations, the propagation of the induced field is modeled in a fully non-linear fashion. 
\end{enumerate}
By construction of this procedure, we can recover the induced magnetic field in Callisto's plasma environment, including its non-linear feedbacks, by subtracting the reference simulation from the full simulation results.
Here, the term \textit{full simulation} refers to a simulation that includes an excited induced magnetic field.
In contrast, with the term \textit{superposition model} we refer to the result of superimposing the induced magnetic field obtained from Equation~(\ref{eq:motivation_Bsec}) on the reference simulation.

In our model, we only investigate the propagation of the induced magnetic field and prescribe the induction process itself.
Thus, all asymmetries in the simulated induced magnetic field can be attributed solely to the propagation.
The inner boundary of our simulation domain describes the boundary layer between the region of origin of the induced magnetic field and the plasma.
It therefore may be representative of Callisto's icy surface or the top of a thin, radially symmetric ionospheric shell, although strictly speaking, this ionospheric layer would already be part of the plasma environment.

We introduce our MHD model in Section~\ref{subsec:model_mhd-model}.
Similar models have already been successfully used to model the space plasma environments of Ganymede \cite{Duling2014,Duling2022} and Europa \cite{Cervantes_2022}.
In Section~\ref{subsec:model_values-solver}, we provide our choice of model values, which we motivate in \ref{appx:values}.
Our model is applied in Callisto's frame of reference.
We use a Cartesian as well as a spherical coordinate system, both with origin in Callisto's geometric center.
In the Cartesian system, the $z$-axis is parallel to Jupiter's rotation axis, the $y$-axis points towards Jupiter's center and the $x$-axis, which completes the right-handed coordinate system, is roughly aligned with Callisto's direction of orbital movement.
In the spherical system, the polar angle $\theta$ is taken from the positive $z$-axis, while the azimuthal angle $\phi$ is counted from the positive $y$-axis (the Jupiter facing meridian) in an easterly direction.

\subsection{MHD Model}
\label{subsec:model_mhd-model}

We employ an MHD model that treats the plasma as a singly charged fluid.
A limitation of this fluid treatment is that it cannot resolve asymmetries in the interaction caused by the gyration of the ions.
However, for positions of Callisto outside the current sheet, the gyroradii of the bulk plasma are only on the order of kilometers \cite{Hartkorn2017b}, and the associated asymmetries are small \cite{Liuzzo2015}.
Therefore, and because this study focuses on the general aspects of propagation in a preferably simple setup, we choose an MHD approach.

At the heart of our MHD model are four Equations~(\ref{eq:model_continuity}~--~\ref{eq:model_induction}) that respectively describe the temporal and spatial evolution of the plasma mass density $\rho$, the bulk velocity $\mathbf{v}$, the thermal pressure $p$ and the magnetic flux density $\mathbf{B}$.
\begin{linenomath*}
\begin{eqnarray}
    \label{eq:model_continuity}
        \frac{\partial\rho}{\partial t} 
        + \nabla \cdot \Big[\rho\mathbf{v}\Big] 
    & = &
        P m_n - L m_L  \\
    \label{eq:model_momentum}
        \frac{\partial\rho\mathbf{v}}{\partial t}
        + \nabla \cdot \Big[
            \rho\mathbf{v}\mathbf{v}
            - \frac{1}  {\mu_0}\mathbf{B}\mathbf{B}
            + \mathbf{I} (p + \frac{1}{2}\frac{B^2}{\mu_0})
        \Big] 
    & = &
        - (L m_L  + \nu_n \rho) \mathbf{v} \\
    \label{eq:model_energy}
        \frac{\partial E_t}{\partial t}
        + \nabla \cdot \Big[
            (E_t + p + \frac{1}{2}\frac{B^2}{\mu_0})\mathbf{v}
            - \frac{1}{\mu_0}\mathbf{B}(\mathbf{v}\cdot\mathbf{B})
        \Big]
    & = &
        -\frac{1}{2}(L m_L + \nu_n \rho) v^2  \nonumber \\
    	& - & \frac{3}{2}(L m_L + \nu_n\rho)\frac{p}{\rho}  \nonumber \\
    	& + & \frac{3}{2}(P m_n + \nu_n \rho) \frac{k_B T_n}{m_n}  \\
    \label{eq:model_induction}
        \frac{\partial\mathbf{B}}{\partial t} 
        - \nabla \times \Big[\mathbf{v} \times \mathbf{B}\Big]
    & = & 
        0
\end{eqnarray}
\end{linenomath*}
In Equation~(\ref{eq:model_energy}), $E_t$ is the total energy $E_t = \frac{3}{2}p + \frac{1}{2} \rho v^2 + \frac{1}{2} \frac{B^2}{\mu_0}$.
Essentially, the left side of the equations is a conservative formulation of MHD, while the right side accounts for the interactions between the magnetospheric plasma and Callisto's atmosphere.
In our model, we include ion-neutral collisions with an effective collision frequency $\nu_{n}$ and ionization and recombination with a production rate $P$ and loss rate $L$, respectively.
The remaining parameters represent the average mass of neutral particles $m_n$ and plasma ions $m_L$, as well as the temperature of the atmosphere $T_n$. $\mathbf{I}$ denotes the identity matrix.

For the sake of simplicity, we describe Callisto's atmosphere as a single species, radially symmetric O\textsubscript{2} atmosphere in hydrostatic equilibrium, neglecting any day/night or ram/wake side asymmetries.
The number density $n_n$ is given by
\begin{linenomath*}
\begin{equation}
    \label{eq:model_atmosphere}
    n_n \left( \mathbf{r} \right)
    = \frac{N_n}{H} \cdot \exp \left( \frac{R_\mathrm{C} - r}{H} \right),
\end{equation}
\end{linenomath*}
with the column density $N_n$, the scale height $H$ and Callisto's radius $R_\mathrm{C} = 2410$\,km.

Due to bulk flow and stochastic thermal motions, the populations of plasma ions and atmospheric particles collide and thereby exchange momentum and energy.
In our model, these processes are controlled by an effective collision frequency $\nu_{n}$, which we parameterize based on the ion-neutral collision cross section $\sigma$ and a typical velocity $v_0$:
\begin{linenomath*}
\begin{equation}
    \label{eq:model_collision-frequency}
    \nu_{n} \left(\mathbf{r} \right)
	= \sigma v_0 n_n \left( \mathbf{r} \right) .
\end{equation}
\end{linenomath*}
For the collision cross section, we adopt a value of $\sigma = 1.5 \cdot 10^{-19}\,$m\textsuperscript{2}, similar to that of \citeA{Duling2022}.
   
The ionization of atmospheric particles as well as recombination enter our model with a production rate $P$ and loss rate $L$, respectively:
\begin{linenomath*}
\begin{equation}
    \label{eq:model_production}
    P\left(\mathbf{r}\right)
	= \nu_\mathrm{ion} \left(\mathbf{r}\right) n_n \left( \mathbf{r} \right) \,,
\end{equation}
\end{linenomath*}
\begin{linenomath*}
\begin{equation}
    \label{eq:model_recombination}
    L = \frac{\alpha}{m_L^2} \left( \rho \left( \mathbf{r}\right) - \rho_0 \right)^2  \qquad \mathrm{if}\,\, \rho \left(\mathbf{r}\right) > \rho_0 ,
\end{equation}
\end{linenomath*}
were $\nu_\mathrm{ion}$ is the ionization frequency.
Since we assume the atmosphere to be at rest in Callisto's frame of reference, i.e., to be momentumless, the production does not enter our momentum equation.
In our model, recombination is only active where the plasma density is higher than the upstream value $\rho_0$, and is controlled by a recombination rate coefficient $\alpha = 2 \cdot 10^{-13}\,$m\textsuperscript{3}\,s\textsuperscript{-1}.

The induced magnetic field enters our model as a prescribed dipole field at the inner boundary, given by Equation~(\ref{eq:motivation_Bsec}).
Only at this boundary, we explicitly describe the induced magnetic field using the boundary conditions derived by \citeA{Duling2014}.

\subsection{Model Parameters and Numerical Solver} 
\label{subsec:model_values-solver}

We apply our model to two scenarios: to a simplified scenario with symmetric geometry and to a case study of the Galileo C03 and C09 flybys.
Since both flybys occurred when Callisto was well above or below the plasma sheet, our simplified model aims to reflect these off-plasma sheet conditions too.
In Table~\ref{tab:model_values}, we provide our model values.
A motivation for this particular choice of values is included in Appendices~\ref{subappx:values_plasma}~and~\ref{subappx:values_callisto}.
The values can be divided into two categories: values that describe the upstream magnetospheric plasma (first section of Table~\ref{tab:model_values}) and values that characterize Callisto (second section).
Due to the limited and often indirect measurements, most of these values are not well constrained.

\begin{table}
	\caption{
		Summary of initial values for our model setups.$^{a}$
    }
    \label{tab:model_values}
	\centering
	\begin{tabular}{l l c c c}
		 \hline
		 Parameter		& Unit              & Symmetric Model & C03                	& C09               \\
		 \hline
		 $\rho$         & amu$\,$m$^{-3}$   & 0.96        &                         &                   \\
		 $\textbf{v}$   & km$\,$s$^{-1}$    & (192, 0, 0) &                         &                   \\
		 $\textbf{B}$   & nT 			    & (0, 35, 0)  & (-4.2, -31.8, -11.5)    & (3.9, 36.9, -9.7) \\
		 $p$            & nPa 			    &  $9.6 \cdot 10^{-2}$ &                &                   \\
         \hline
         $N_n$                & cm$^2$      & $2.1\cdot 10^{15}$   &                &                   \\
         $H$                  & km          & 230         & 60                      & 60                \\
         $m_n$                & amu         & 32          &                         &                   \\
         $T_n$                & K           & 300         &                         &                   \\
         $\nu_\mathrm{ion}$ & s$^{-1}$    & $1.2\cdot 10^{-9}$   & $3.0\cdot 10^{-9}$, $3.0 \cdot 10^{-8}$  & $3.0\cdot 10^{-9}$, $3.0 \cdot 10^{-8}$  \\
		 \hline
		 $v_a$          & km$\,$s$^{-1}$    & 782         & 761                     & 857               \\
		 $c_s$          & km$\,$s$^{-1}$    & 317         & 	                    & 	                \\
		 $v_{f,\perp}$  & km$\,$s$^{-1}$    & 844         & 825                     & 914               \\
		 \hline
		 \multicolumn{5}{l}{$^{a}$For the flyby setups, only the values that differ from the symmetric model are given.} \\
		 \multicolumn{5}{l}{The two values for $\nu_\mathrm{ion}$ correspond to the shadow and sunlit side, respectively.}
	\end{tabular}
\end{table}

We carry out our numerical simulations with PLUTO, a finite-volume based code specialized in modeling astrophysical fluids \cite{Mignone2007}.
Our version of the code incorporates suitable boundary conditions for modelling an icy moon that were derived by \citeA{Duling2014}.
These conditions treat Callisto as having a non-conducting, plasma-absorbing surface and allow to include an induced magnetic field in the simulation process.
On the upstream side of the outer boundary ($\phi \leq 180^\circ$), we use fixed boundary conditions that are representative of the inflowing plasma, while on the downstream side ($\phi > 180^\circ$), we employ open conditions.
Distributing ${\sim}8\cdot 10^6$ and ${\sim}11 \cdot 10^6$ cells non-equidistantly in the spherical simulation domain between $1 -100\,R_\mathrm{C}$, we achieve maximal radial resolutions at Callisto's surface of $0.02\,R_\mathrm{C}$ and $0.008\,R_\mathrm{C}$ for the simplified scenario and the flyby scenarios, respectively.
A more detailed description of the simulation setup is included in Appendix \ref{subappx:values_numerics}.


\section{Results: Symmetric Model}
\label{sec:results}

We investigate the propagation of the magnetic field perturbation due to an induced magnetic field on the basis of a reference simulation.
Thus, this reference simulation sets the physical conditions of the propagation and is presented in Section~\ref{subsec:results_overview}.
Subsequently, we analyze the propagation's spatial and temporal implications on the induced magnetic field in Sections \ref{subsec:results_spatial} and \ref{subsec:results_temporal}.

\subsection{Overview of Callisto's Local Plasma Environment}
\label{subsec:results_overview}

\begin{figure}
	\noindent\includegraphics[width=\textwidth]{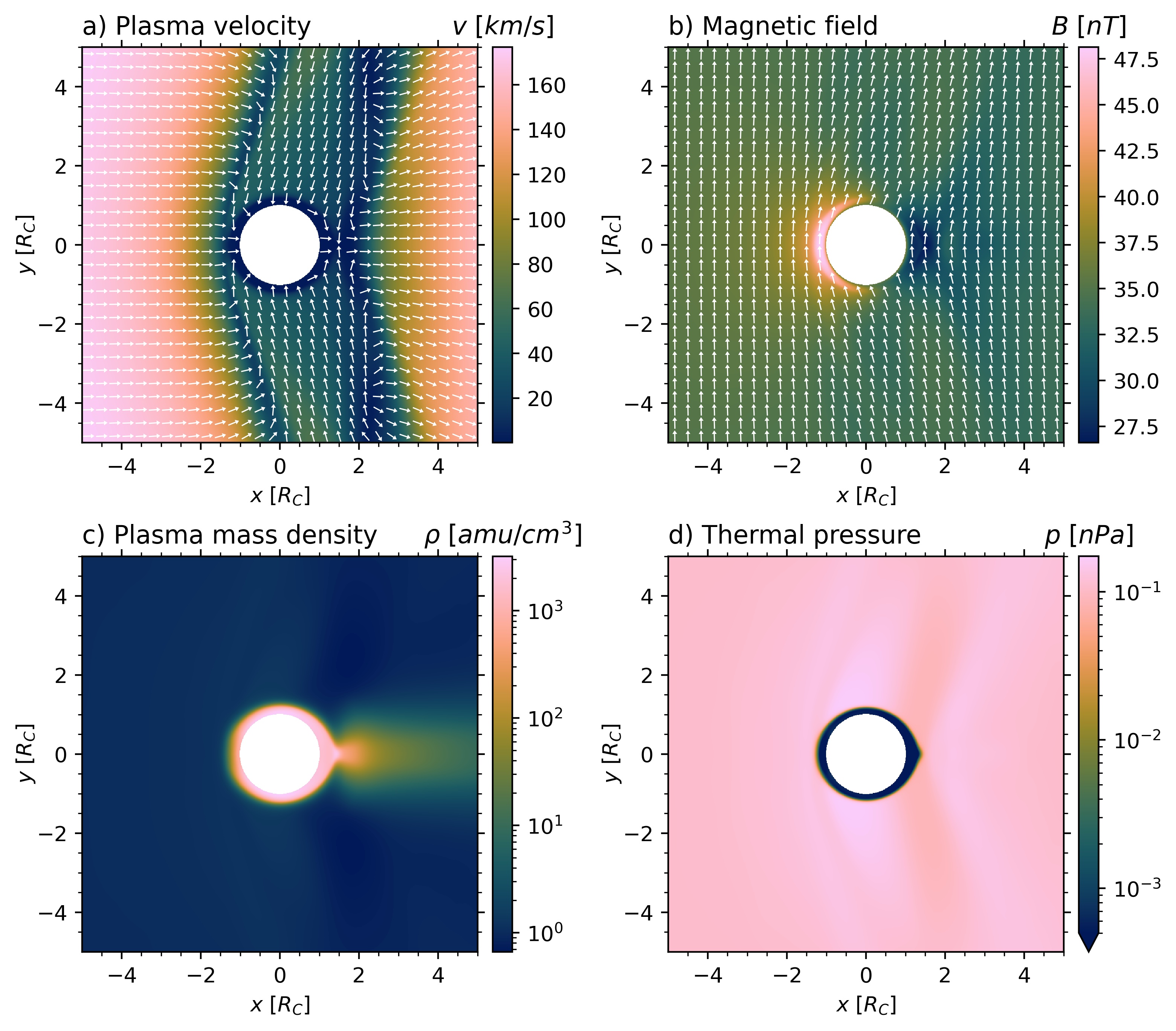}
	\caption{
		Model variables of the reference simulation in the $xy$-plane. The panels show (a) the plasma bulk velocity, (b) the total magnetic field, (c) the plasma mass density and (d) the thermal pressure.
		For the two vector variables, the white arrows indicate the projection of the direction onto the plane.
	}
	\label{fig:results_overview-simulation}
\end{figure}

Figure~\ref{fig:results_overview-simulation} shows the modelled MHD variables of the reference simulation in the $xy$-plane.
Close to Callisto, within a few scale heights of the atmosphere, ionization leads to a dense plasma with plasma mass densities roughly three orders of magnitude above the background value $\rho_0$.
Ion-neutral collisions drain kinetic and thermal energy from the magnetospheric plasma. 
Near Callisto, the thermal pressure is reduced by about two orders of magnitude.
Within a few scale heights of the atmosphere, the plasma velocity is decreased to a few km/s.
While this velocity perturbation is continued into the Alfvén wings, it is partly compensated by an acceleration toward Callisto caused by the pressure gradient.
The magnetic field is shaped by moon-magnetosphere interactions: its magnitude is enhanced on the upstream side within the so-called pile-up region and decreased on the downstream side, where reacceleration occurs.
Within the Alfvén wings, the direction of the magnetic field and of the bulk motion are \mbox{(anti-)}\hspace{0pt}parallel. 

In terms of plasma density, our model predicts that the local environment is strongly dominated by freshly ionized plasma, consistent with plasma wave observations by \citeA{Gurnett1997,Gurnett2000}.
In our radially symmetric atmosphere, 95\% of the ions are produced within a distance of 3H, where mean bulk and thermal velocities are low.
Consequently, the bulk of the picked-up plasma acquires gyroradii of $r_{g} <30\,$km, significantly smaller than $R_\mathrm{C}$ and thereby justifying our MHD approach.

\begin{figure}
	\noindent\includegraphics[width=\textwidth]{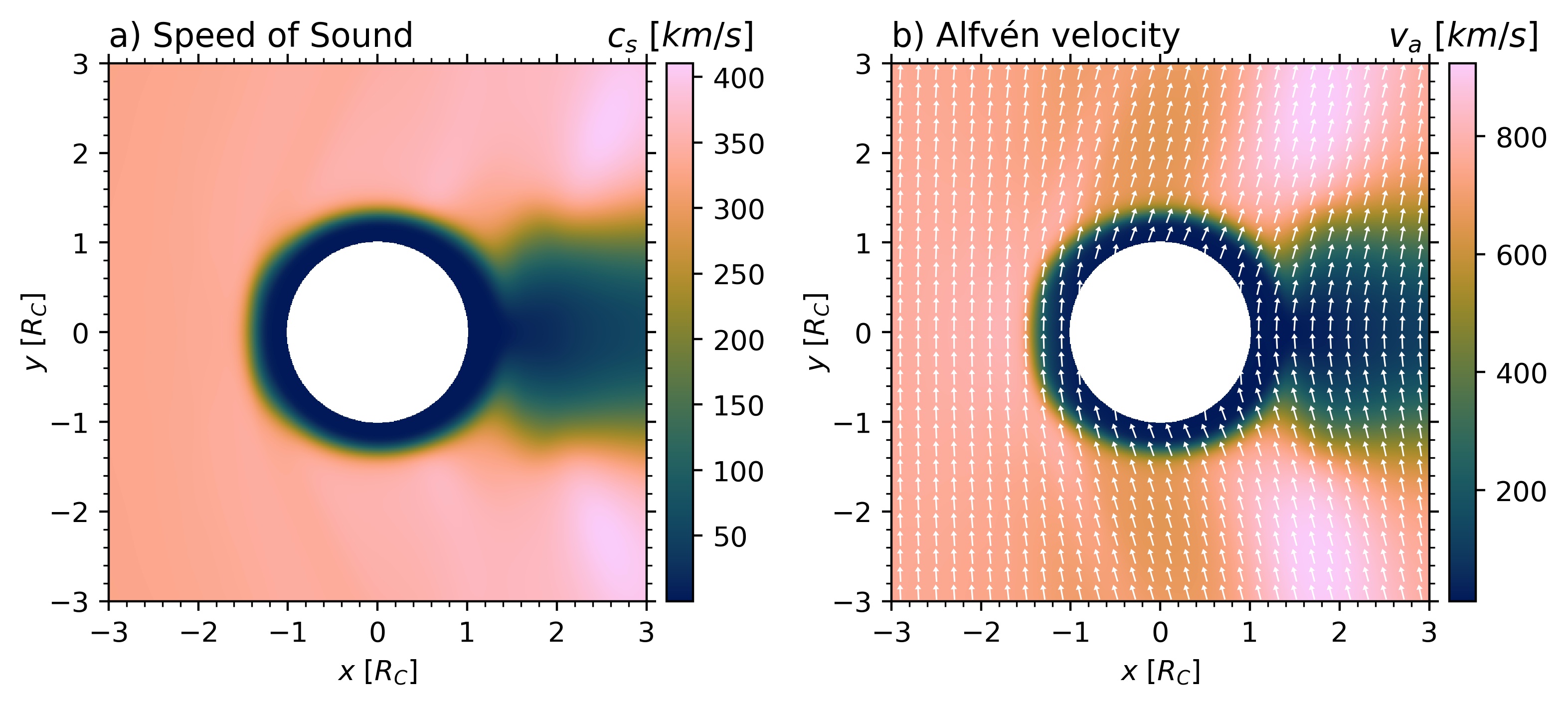}
	\caption{
		Speed of sound $c_s$ (left) and Alfvén velocity $v_a$ (right) in the $xy$-plane.
		In case of the Alfvén velocity, the white arrows show the projection of the magnetic field's direction onto the plane.
	}
	\label{fig:results_mode-velocities}
\end{figure}

Figure~\ref{fig:results_mode-velocities} shows the speed of sound $c_s$ (panel~a) and the Alfvén velocity $v_a$ (panel~b).
Both, the speed of sound and the Alfvén velocity, are inversely proportional to the square root of the plasma mass density:
\begin{linenomath*}
\begin{eqnarray}
	\label{eq:results_mode-velocities}
	c_s 
	   = \frac{1}{\sqrt{\rho}} \cdot \sqrt{ \frac{5}{3} p}
	\,, \quad
	v_a
	   = \frac{1}{\sqrt{\rho}} \cdot \frac{B}{\sqrt{\mu_0}} 
	\,.
\end{eqnarray}
\end{linenomath*}
Thus, close to the Callisto, in the area of strongly enhanced plasma mass densities, their magnitudes reduce by more than one order of magnitude.
Additionally, reduced velocities are visible in the approximate geometric wake of Callisto.
In the near-field, i.e., close to the source of the magnetic field perturbation at Callisto's surface, in the region of frequent ion-neutral collisions, the MHD-modes mix and are not clearly separable.  
However, using the velocities in Figure~\ref{fig:results_mode-velocities} as a rough indication, it is evident that (a) the wave velocities are not homogeneous and (b) they strongly reduce in Callisto's local plasma environment.
In the simple geometry of our reference simulation, roughly three regions of interest may be identified.
First, the region upstream of Callisto ($x < -1\,R_\mathrm{C}$) which is located perpendicular to the magnetic field and accessible only to the fast mode.
Second, the equivalent region on the downstream side ($x > 1\,R_\mathrm{C}$), which, in addition to the fast mode, is also accessible to the convection mode.
And third, the regions that are in direction parallel to the magnetic field.
Here, Alfvén waves as well as slow mode waves may additionally transport the induced magnetic field.

\subsection{The Spatial Implications}
\label{subsec:results_spatial}

\begin{figure}
	\noindent\includegraphics[width=\textwidth]{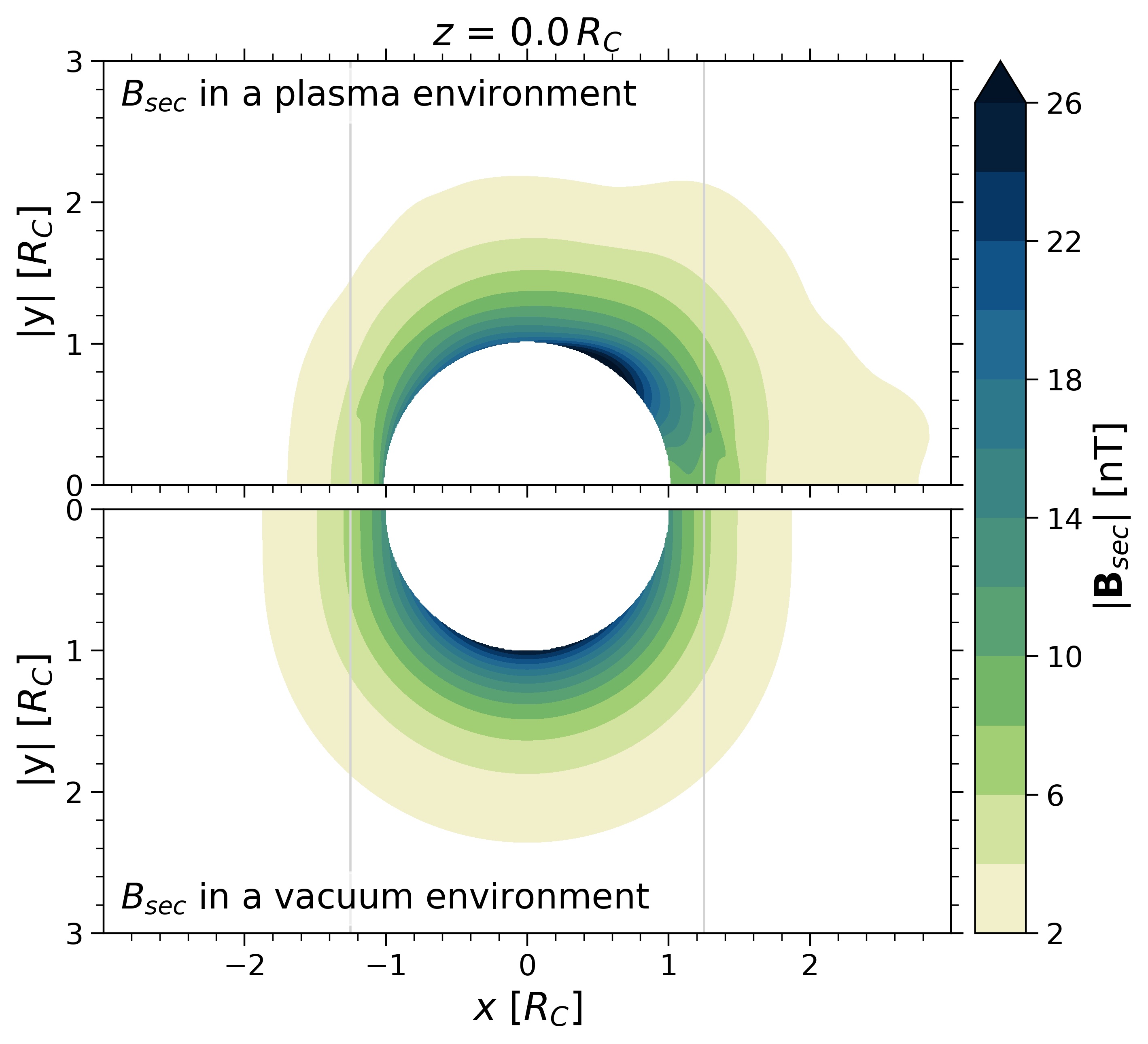}
    \caption{
		Contour plot of the magnitude of the induced magnetic field in the $xy$-plane.
		The upper half shows the induced magnetic field in the plasma environment from the MHD simulations (Equation~(\ref{eq:results_Bsec-simulation})), the lower half shows the induced magnetic field in a vacuum environment (Equation~(\ref{eq:motivation_Bsec})).
		The thin gray lines indicate the up- and downstream profile, respectively.
		Our simplified model is symmetric to the $y=0$-plane, allowing for this mirrored representation.
	}
	\label{fig:results_contour}
\end{figure}

Figure~\ref{fig:results_contour} compares the magnitude of the induced magnetic field $B_\mathrm{sec}$ due to the transport effects in a plasma with that in a vacuum.
For this comparison, the $xy$-plane is selected as it includes the $\mathbf{v}_0$ and $\mathbf{B}_0$ vectors.
Per construction of our model, we identify the induced magnetic field propagated in the plasma environment $\mathbf{B}_\mathrm{sec}^{P}$,including its non-linear feedbacks, as the difference between the reference and the full simulation:
\begin{linenomath*}
\begin{equation}
	\label{eq:results_Bsec-simulation}
	\mathbf{B}_\mathrm{sec}^P
		= \mathbf{\Delta B}
		= \mathbf{B}_\mathrm{full} - \mathbf{B}_\mathrm{ref} \,.
\end{equation}
\end{linenomath*}
The induced magnetic field in a vacuum environment $\mathbf{B}_\mathrm{sec}^V$ is given by Equation~(\ref{eq:motivation_Bsec}).
The spatial properties of the induced magnetic field differ between the plasma and the vacuum.
In a vacuum or non-conducting environment, the induced magnetic field is a plain dipole and symmetric with respect to the dipole axis, which is aligned with the y-axis in our model setup.  
In the magnetospheric plasma environment, however, the induced magnetic field is asymmetric with a distinct upstream/downstream asymmetry.
On the upstream side ($x < 0$), it is compressed, meaning it decays more strongly than with the cubed distance.
On the downstream side ($x > 0$), in contrast, it extends further, i.e., it decays less strongly.  
The areas of the highest induced magnetic field magnitude, which are at the dipole poles in the vacuum environment, are wrapped and convected into the wake of Callisto.
In the vacuum environment, magnetic perturbations $\geq 2$\,nT are observable only up to $x < 2\,R_\mathrm{C}$.
In contrast, in the plasma environment on Callisto's downstream side, such perturbations are observable even at a distance of $x > 2.5\,R_\mathrm{C}$.

\begin{figure}
	\noindent\includegraphics[width=\textwidth]{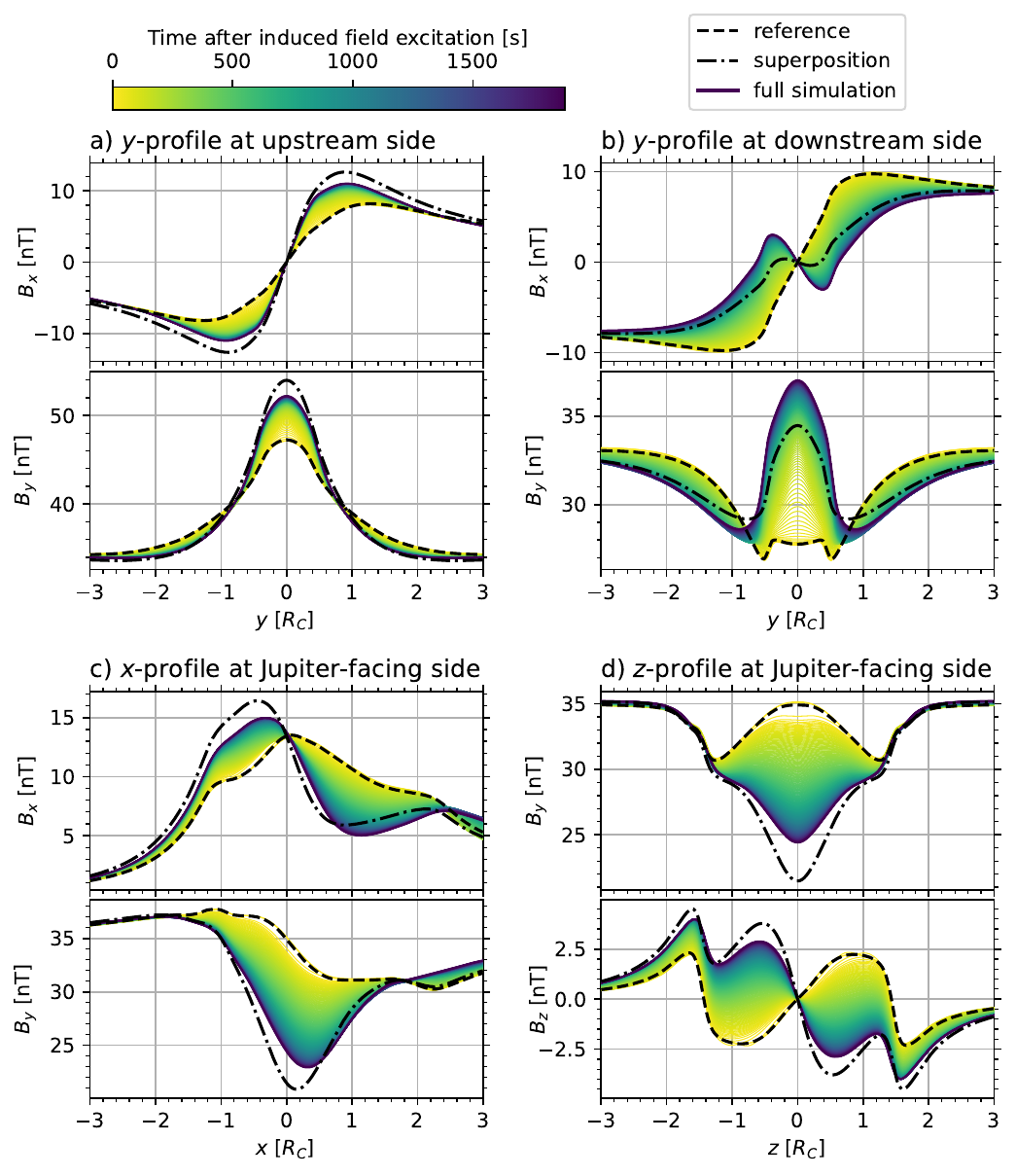}
	\caption{
		Components of the total magnetic field at selected profiles color coded as a function of time.
		These profiles are: (a) $y$-profile at ${x=-1.25\,R_\mathrm{C}}, {z=0}$, (b) $y$-profile at ${x=1.25\,R_\mathrm{C}}, {z=0}$, (c) $x$-profile at ${y=1.25\,R_\mathrm{C}}, {z=0}$, and (d) $z$-profile at ${x=0}, {y=1.25\,R_\mathrm{C}}$.
		The lines indicate: the magnetic field of the reference simulation (dashed black), the superposition of reference simulation and induced magnetic field in a vacuum environment (dash-dotted black), and the magnetic field of the full simulation at continuous time stages after exiting the induced magnetic field (color coded).
		}
	\label{fig:results_profiles}
\end{figure}

The upstream/downstream asymmetry of the induced magnetic field in the plasma environment further manifests in the results visualized in Figure~\ref{fig:results_profiles}.
To gain a quantitative understanding of the transport effect on the induced magnetic field, we compare the total magnetic fields at four profiles: one profile located upstream of Callisto (panel a), one profile located downstream (panel b) and two profiles that are perpendicular to the unperturbed magnetic field (profiles c \& d).
In these plots, the temporal evolution of the full simulation after the excitation of the induced magnetic field is included with the color code.
We will address the temporal aspects in Section~\ref{subsec:results_temporal}.
In this paragraph, we will only discuss the final state of the full simulation (dark blue line), which shows the fully developed induced magnetic field.
The results of the reference simulation (dashed black line) demonstrate that close to Callisto, i.e., at the centers of the profiles, moon-magnetosphere interactions such as pile-up and draping perturb the magnetic field considerably with deviations up to ${\sim}13$\,nT.
The superposition of reference simulation and induced magnetic field in a vacuum is shown as dash-dotted black line.
This superposition model can be compared to the results of the full simulation (dark blue line), which additionally considers the transport effects on the induced magnetic field.
At the upstream profile (panel a), where induction field and moon-magnetosphere interactions have a somewhat similar imprint, the observable magnetic field in the plasma environment (dark blue line) falls short of the magnitudes predicted by the superposition (dash-dotted black line). 
This is in contrast to the downstream profile (panel b).
Here, by considering the propagation effect, the observable imprints of the induced magnetic field are increased compared to the vacuum environment.
These imprints are, for example, the sign reversal of the $B_x$-component at $y \approx \pm 0.5$\,R\textsubscript{C} as well as the enhanced $B_y$-strength at the induced magnetic field's equator at $y=0$, were the induced magnetic field is parallel to the magnetic background field.

In a vacuum environment, the induced magnetic field is strongest at its poles, where it opposes the inducing magnetic field.
Accordingly, the superposition model (dash-dotted black line) predicts the lowest $B_y$ magnitudes at $x \approx 0$ (panel c \& d of Figure \ref{fig:results_profiles}).
In the plasma environment, however, the imprint of the induced magnetic field at the pole axis ($x=0, z=0$) is less pronounced than predicted.
Here, the region of the strongest induced magnetic field and thus the weakest total $B_y$-magnitude shows in the downstream hemisphere (panel c and also Figure~\ref{fig:results_contour}).
In our setup, the pole axis of the prescribed induced magnetic field is antiparallel to the background magnetic field, thus located in the foot of the Alfvén wings.
Inside the Alfvén wings, the magnetic field bends downstream.
Perturbations that propagate (anti-)parallel to the magnetic field are thus displaced downstream, which could contribute to the downstream shift of the $B_y$-minima.

\begin{figure}
	\noindent\includegraphics[width=\textwidth]{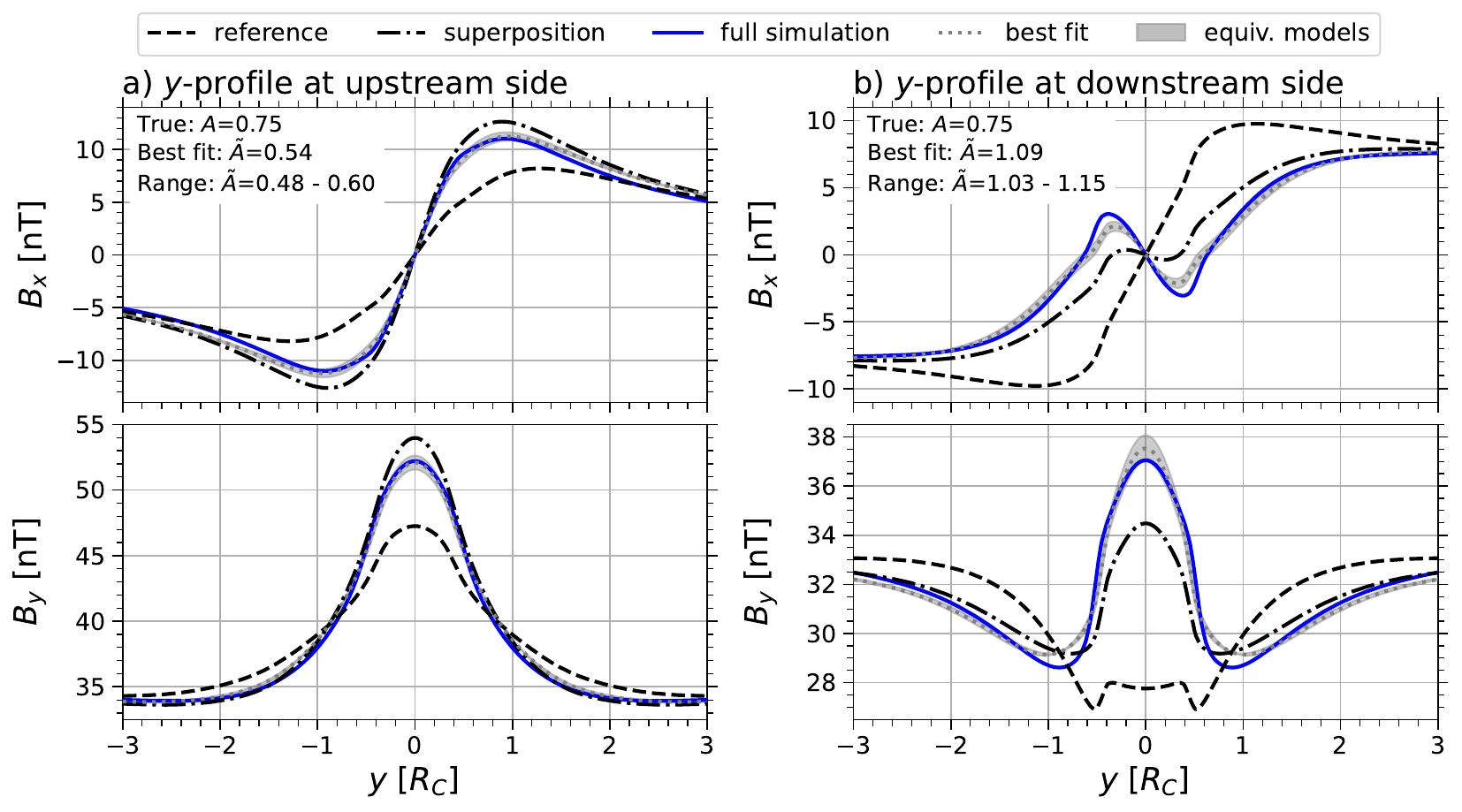}
	\caption{
		Best fit solutions of the superposition model fitted to the full simulation at the upstream profile (panel a, left) and the downstream profile (panel b, right).
		The amplitude $\tilde{A}$ of the best fit and the amplitude range of models that fit similarly well are given in the upper left corners of the plots.
	}
	\label{fig:results_inversion}
\end{figure}

The finding that in a plasma environment, the properties of an induced magnetic field are less pronounced on the upstream side and more pronounced on the downstream side than in a vacuum environment, has an important consequence.
This implies that using the analytical solution derived for the vacuum environment (Equation~(\ref{eq:motivation_Bsec})), the strength of the induced magnetic field is underestimated on the upstream side and overestimated on the downstream side.
We demonstrate this consequence in Figure~\ref{fig:results_inversion}.
Here, we treat the simulation results of the full simulation (at the final state) as observed magnetic field (blue line) and inversely calculate the induced magnetic field of a vacuum environment that fits best to this data (dotted gray line).
For this inverse calculation, we minimize a root-mean-square deviation (RMS) as a function of a plasma-altered amplitude $\tilde{A}$, searching in the interval $0 \leq \tilde{A} \leq 2$:
\begin{linenomath*}
\begin{equation}
	\label{eq:results_rms}
	\mathrm{RMS} \left( \tilde{A} \right)
	= \sqrt{\frac{1}{2N} \sum_{i=1}^N \left\|
			\mathbf{B}_{\mathrm{full}, i}
			- \left(\mathbf{B}_\mathrm{ref} + \tilde{A} \cdot \mathbf{B}_\mathrm{sec}^\infty \right)_i
		\right\|^2} .
\end{equation}
\end{linenomath*}
In this equation, where $N=1500$ is the number of samples along a profile, we set the denominator to $2N$ because the $z$-component of the induced magnetic field vanishes in the $xy$-plane.
While the true amplitude of the induced magnetic field used in the simulation is $A=0.75$, the best fitting amplitudes in terms of the RMS are $\tilde{A}=0.54$ and $\tilde{A}=1.09$ for the upstream and downstream profile, respectively. 
The gray shaded bands in Figure~\ref{fig:results_inversion} depict models that fit similarly well to the observable magnetic field, when allowing for an arbitrarily chosen  $15\,\%$ increase in the RMS value.
The resulting ranges for the observed amplitude $\tilde{A}$ are $0.48 - 0.60$ and $1.03 - 1.15$ for the upstream and downstream side, respectively.
Thus, using the analytic solution derived for a vacuum environment, the strength of the induced magnetic field is underestimated by $20 - 36$\,\% on the upstream side and overestimated by $37 - 53$\,\% on the downstream side.

\begin{figure}
    \noindent\includegraphics[width=\textwidth]{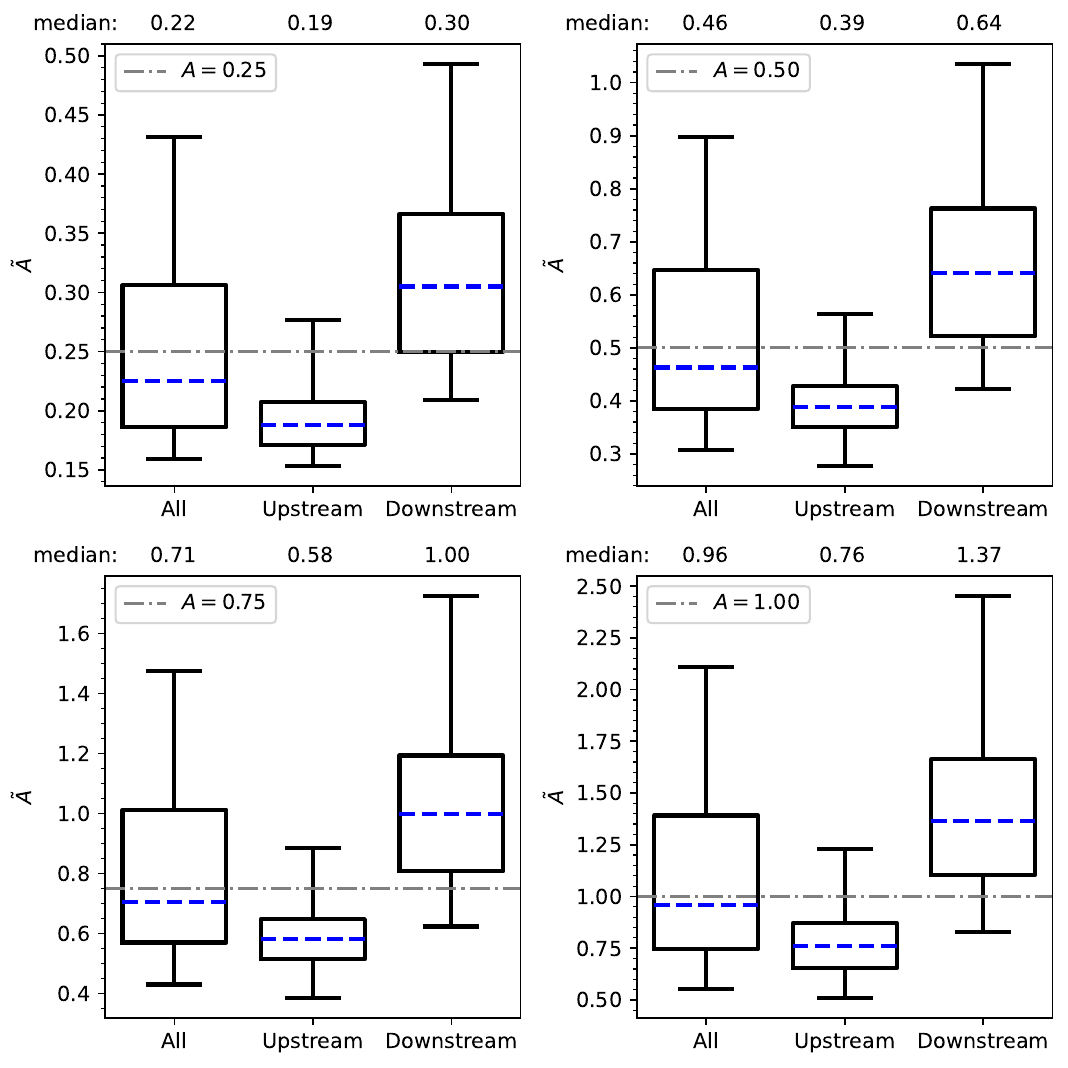}
    \caption{
		Distributions of the transport-altered amplitude $\tilde{A}$ based on Equation~(\ref{eq:results_A-from-ratio}).
		The main body of the box plots represents the data in the 25-75 percentile range, the ends of the whiskers show the 5 and 95 percentile, respectively.
		The respective median value is indicated by the dashed blue line and given at the top of each panel. 
		The dash-dotted gray line displays the true amplitude used in the simulations.
	}
	\label{fig:results_boxplots}
\end{figure}

The results for the two profiles are limited to the two respective spatial domains.
Thus, for a more comprehensive view, we examine the induced magnetic fields within the entire spherical domain $r < 4\,R_\mathrm{C}$.
Assuming that the induced magnetic field within a plasma can be represented similarly to the vacuum case (Equation~(\ref{eq:motivation_Bsec})) but with a plasma-altered induction amplitude $\tilde{A}$, we define:
\begin{linenomath*}
\begin{equation}
	\label{eq:results_A-from-ratio}
	\tilde{A} \left(\mathbf{r}\right)
	= \frac{B_\mathrm{sec}^P \left(\mathbf{r}\right)}{B_\mathrm{sec}^\infty \left(\mathbf{r}\right)} .
\end{equation}
\end{linenomath*}
We apply Equation~(\ref{eq:results_A-from-ratio}) to sample locations consisting of 60 equally spaced spherical shells, on each of which we distribute 2500 points equidistantly.
We omit sample locations where the magnitude of the induced fields is small ($A \cdot B_\mathrm{sec}^\infty <1$nT) or where numerical variability in the reference simulation leads to uncertainties $\delta \tilde{A} > 10\,$\%.
Figure~\ref{fig:results_boxplots} shows the resulting amplitudes $\tilde{A}$ for four simulation setups.
All four setups, which employed a true amplitude of 0.25, 0.50, 0.75 and 1.00, exhibit a similar pattern and confirm the results shown in Figure~\ref{fig:results_inversion}: there are clear differences between the upstream hemisphere ($x < 0$) and the downstream hemisphere ($x > 0$).
In the upstream hemisphere (middle box plots), more than 75\,\% of the obtained amplitudes $\tilde{A}$ fall below the true value.
Judged by the median, the strength of the induced magnetic field is underestimated by at least 22\,\%.
In the downstream hemisphere (right box plots), in contrast, 75\,\%  or more of the obtained amplitudes $\tilde{A}$ are above the true value.
The strength of the induced magnetic field is overestimated here by at least 20\,\%.
Over the entire volume (left box plots), the computed medians $\tilde{A}$ are slightly below the true amplitudes.
This reduction can be explained by the anisotropic propagation properties of the MHD modes: The spatial domain accessible to the convection mode is smaller compared to the spatial domain that remains inaccessible.
Additionally, ion-neutral collisions act to dampen the amplitudes of the MHD wave modes due to induction and the magnetic perturbations transported by them.
The range between the 5\,\% and 95\,\% percentile as well as between the 25\,\% and 75\,\% percentile is smaller in the upstream than in the downstream hemisphere, i.e., the transport effects on the induced magnetic field are more similar.
This feature possibly is related to the fact that propagation in parts of the upstream hemisphere is limited to the more isotropic fast mode, whereas the advection downstream is more strongly spatially structured.

\begin{figure}
	\noindent\includegraphics[width=\textwidth]{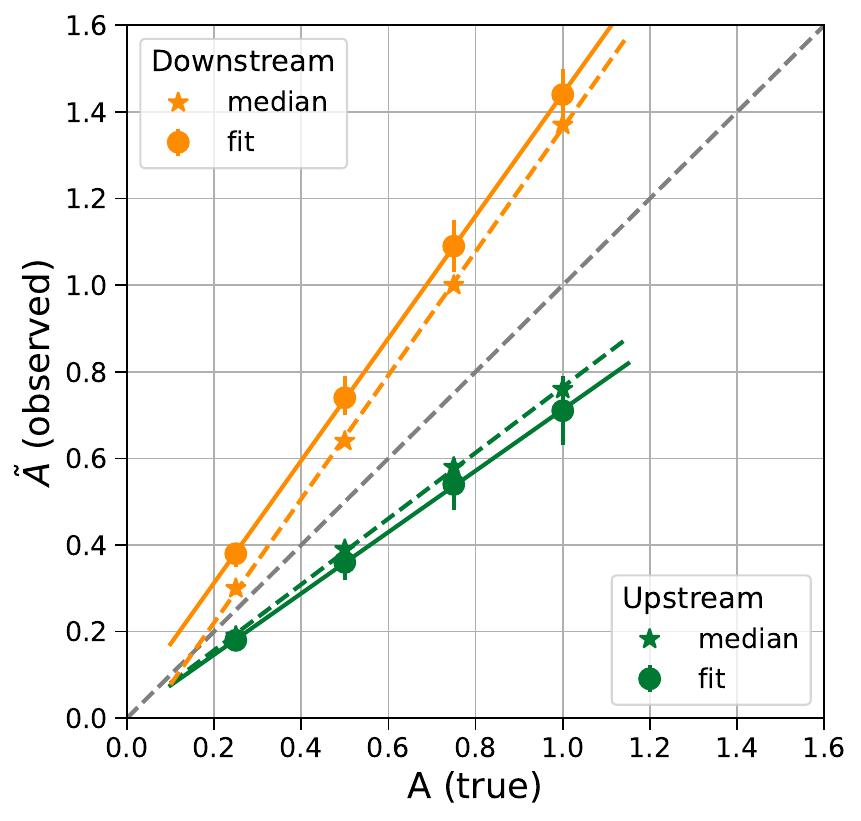}
	\caption{
		Comparison of true ($A$) and transport-altered ($\tilde{A}$) amplitudes for both methods, the RMS minimization (dots) and the median based estimate (stars).
		The upstream profile/hemisphere is plotted in green, the downstream profile/hemisphere in orange.
		The respective solid and dashed lines indicate least-squares linear regressions.
	}
	\label{fig:results_observed-vs-true}
\end{figure}

The described effect of over/underestimating the strength of the induced magnetic field is, in relative terms, approximately independent of the true amplitude $A$.
This is visualized in Figure~\ref{fig:results_observed-vs-true}, where we show the plasma-altered amplitude $\tilde{A}$ as a function of the true amplitude $A$ for both methods, the RMS minimization (dots) and the median based estimate (stars).
At both sides, upstream (green) and downstream (orange), and for both methods, the plasma-altered amplitude $\tilde{A}$ exhibits a linear dependency on the true amplitude $A$.
The ratio of plasma-altered to true amplitude $\tilde{A}/A$ is, on average, 0.74 on the upstream side and 1.39 on the downstream side.

\subsection{The Temporal Implications}
\label{subsec:results_temporal}

In Callisto's plasma environment, the MHD modes — and being transported by them the induced magnetic field — propagate at velocities that can reduce to a few km\,s\textsuperscript{-1} (see Figure~\ref{fig:results_mode-velocities}).
This results in a temporal delay between the emergence of the induced magnetic field at the surface of the moon and its arrival at spacecraft location.
This temporal delay manifests as an additional phase shift on top of the phase shift caused by the induction process in a medium with a finite conductivity, such as a subsurface ocean or an ionosphere.
In this propagation process, two separate times need to be investigated:
\begin{itemize}
	\item[(i)] The time of the arrival of the wave front from the surface.
	\item[(ii)] The time required for the wave pattern to fully develop, i.e., for a standing wave to establish. 
\end{itemize}

\begin{figure}
	\noindent\includegraphics[width=\textwidth]{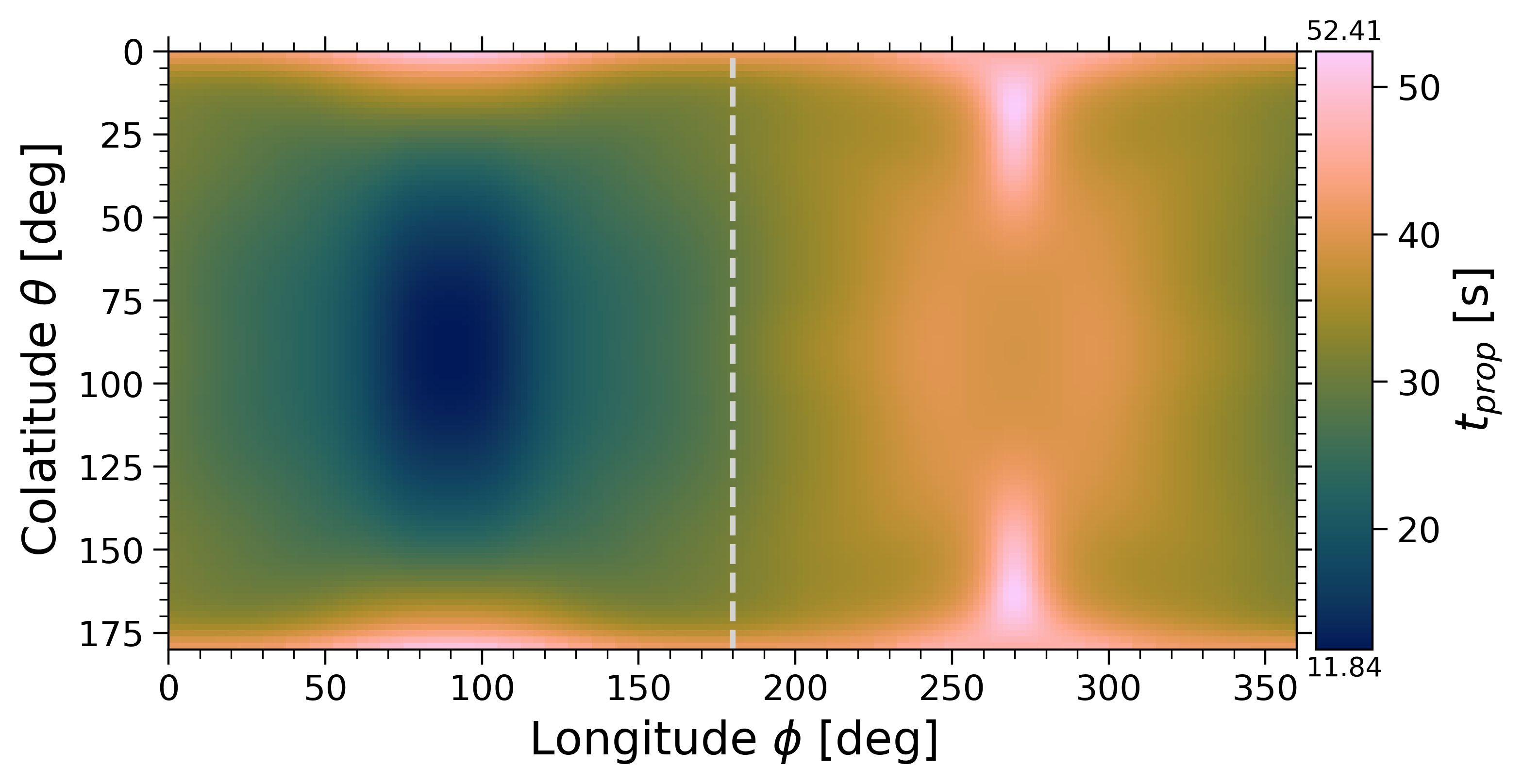}
	\caption{
		Estimate of the induced magnetic field's minimal propagation time to reach the spherical shell at $r = 1.25\,R_\mathrm{C}$.
		The dashed gray line indicates the anti-Jovian meridian. The left half ($0-180$\,degrees) shows the upstream hemisphere, the right half ($180-360$\,degrees) show the downstream hemisphere.
	}
	\label{fig:results_propagation-time}
\end{figure}

Figure~\ref{fig:results_propagation-time} shows the minimal time (i) that is required for the induced field to reach the spherical shell at $r = 1.25\,R_\mathrm{C}$, i.e., to cover the distance of $0.25\,R_\mathrm{C}$.
We estimate this minimal time by integrating the slowness ($= 1/v$) of the fastest MHD mode, the fast mode, traveling along the shortest path, the radial direction.
\begin{linenomath*}
\begin{equation}
	\label{eq:results_propagation-time}
	t_\mathrm{prop} \left(\mathbf{r}\right)
	= \int_{R_\mathrm{C}}^r \frac{1}{v_{f, \perp}\left(\mathbf{r^\prime}\right)} dr^\prime
\end{equation}
\end{linenomath*}
The velocity of the fast mode is anisotropic and largest perpendicular to the magnetic field.
Its upper limit, therefore, is given by $v_{f} \leq v_{f, \perp} = \sqrt{c_s^2 + v_a^2}$.
Thus, the wave propagation times color-coded in Figure~\ref{fig:results_propagation-time} represent a lower limit.
The required times range between ${\sim}11$\,s and ${\sim}53$\,s, with a mean value of $\overline{t}_\mathrm{prop} = 31.6 \pm 7.9$\,s. 
The minimum propagation time of $11.84$\,s occurs at the central upstream point ($\theta=90^\circ, \phi=90^\circ$), while at the equivalent point on the downstream side ($\theta=90^\circ, \phi=270^\circ$), with ${\sim}39.1$\,s the propagation time is about three times as high.
These differences in the propagation time are explained by the fact that the velocity of the fast mode is inversely proportional to plasma density $\sqrt{\rho}$ and roughly proportional to the magnitude of the magnetic field $B$, since in our case $v_a > c_s$.
On the downstream side, in Callisto's wake, the plasma mass densities are enhanced (see Figure~\ref{fig:results_overview-simulation}).
Additionally, moon-magnetosphere cause an increase in magnetic field magnitude on the upstream side and a decrease on the downstream side.
For the same reason, comparable low magnetic field strength and high plasma densities, we obtain propagation times $> 40$\,s at Callisto's flanks at $\theta \approx 0^\circ$ or $\theta \approx 180^\circ$.
While the mean propagation time of $31.6\,$s is more than one order of magnitude larger than the wave propagation time in an unperturbed plasma $t_\mathrm{prop} = \linebreak[4]{} {0.25 \cdot R_\mathrm{C}/v_{f, \perp, 0}} = 0.71\,$s (see Table~\ref{tab:model_values}) and several orders of magnitude larger than the propagation time of the ordinary electromagnetic mode $t_\mathrm{prop} \approx 0.25 \cdot R_\mathrm{C}/c = 2 \cdot 10^{-3}\,$s, it is still small compared to the period of the inducing field $T_\mathrm{prim}=10.18$\,h.
Accordingly, the additional phase shift $\phi^\mathrm{ph}_\mathrm{prop} = \overline{t}_\mathrm{prop}/T_\mathrm{prim} \cdot 360^\circ= 0.3^\circ$ is negligible in most applications of the induction method.

The times required until the wave patterns generated at the surface develop into standing waves (ii) is shown in Figure~\ref{fig:results_profiles}, where we plot our simulation results for a continuous selection of time steps after the excitation of the induced magnetic field.
As is visible, the times that are required for the induced magnetic field to fully develop, i.e., to reach a stationary state, are on the order of several minutes rather than on the order of seconds.
At the upstream profile (panel a), were the propagation is limited to the fast mode, the stationary state is attained quickest.
At the downstream profile (panel b) or at the two profiles (panels c \& d) centered at Callisto's sub-Jovian side (the side normal to the direction of the background magnetic field), a longer time span is required for the stationary state.
Here, the induced magnetic field is possibly also propagated by the slow and Alfvén modes as well as the convection mode.
Ion-neutral collisions and mass loading strongly reduce the convection velocities in Callisto's dense atmosphere.
Consistent with hybrid models \cite{Liuzzo2015,Liuzzo2016}, our simulation predicts the plasma bulk velocity to reduce by nearly two orders of magnitude in Callisto's close vicinity.
Within on scale height, we compute an average plasma bulk velocity of $v = 0.23$\,km\,s\textsuperscript{-1} that is close to the value ($v \approx 0.1$\,km\,s\textsuperscript{-1}) estimated by \citeA{Strobel2002}, who accordingly derive a convection time of $R_\mathrm{C}/v \approx 2\cdot 10^4$\,s.
Integrating the slowness of the convection mode up to a distance of $r = 1.25\,R_\mathrm{C}$ in the central wake of Callisto (analogous to Equation~(\ref{eq:results_propagation-time})), yields a convection mode propagation time of $t = 2.7\cdot 10^3$\,s. 
Thus, any perturbation that is transported by the convection mode  will be strongly delayed in Callisto's vicinity.

At an induction period of $T_\mathrm{prim}=10.18$\,h, a phase shift of $\phi^\mathrm{ph} = 1^\circ$ corresponds to a time span of about $102\,$s.
Accordingly, the results in Figure~\ref{fig:results_profiles} indicate additional phase shifts that can exceed ${\phi^\mathrm{ph} > 10^\circ}$.
We note, however, that in our study we excite the induced magnetic field as one full-strength pulse to facilitate a clear characterization of its effects.
At Callisto, meanwhile, the amplitude of the induced magnetic field increases and decreases continuously over the time span $T_\mathrm{prim}/2 \approx 5$\,hours.
Thus, the interference pattern of induced fields continuously adapts rather than builds up, but is still subject to the temporal delays analyzed here.

The effect of a phase shift $\phi^\mathrm{ph} > 0$ is that an induced magnetic field is measured at the position of the space probe that was induced by an earlier primary field.
Depending on the magnitude of the phase shift, the strength (and direction) of the former primary field can differ substantially from its current value.
One source of the phase shift is the induction process itself.
Added to this is the spatially dependent phase shift resulting from the transport of the induced field.
On Callisto's downstream side, where transport time scales can get large, it is thus possible that induced magnetic fields are observable at times, when the primary field is weak, and, vice versa, that the observed induction signals are weak despite a concurrently strong primary field.
Approximating the average rate of change of the primary magnetic field by $\dot{B}_\mathrm{prim} \approx 4 |B_0| \cdot T_\mathrm{prim}^{-1}$, a phase shift of $\phi^\mathrm{ph} = 10^\circ$ translates into a difference of $\Delta B_\mathrm{prim} = 1/9 B_0 \approx 4\,$nT in the primary magnetic field.
Thus, at locations that see a strongly delayed induced magnetic field, the transport phase shift alters the strength of the induction signals by \linebreak[4]{}${\Delta B_\mathrm{sec} \propto A \cdot \Delta B_\mathrm{prim}}$.
This difference in magnitude, which is an effect of the finite velocity of the MHD modes, is in addition to the spacial transport effects presented in Section~\ref{subsec:results_spatial}.

\section{Case Study of Galileo's C03 and C09 Flyby}
\label{sec:galileo}

In this section, we abandon the simplified symmetry and present modelling results for Galileo's C03 and C09 flybys.
Both flybys show induced magnetic fields \cite<e.g.>[]{Zimmer2000}, with C03 being a downstream flyby and C09 an upstream flyby.
For reference, Figure~\ref{fig:trajectories_trajectories} in \ref{appx:trajectories} depicts the trajectories of the two flybys.
To allow a more representative comparison with the measurements, we improve the atmospheric scale height $H$ to 60\,km and consider the directional dependence of the photoionization.
While the near-surface scale height of Callisto's O\textsubscript{2} atmosphere is ${\sim}30\,$km, $H=60$\,km represents a balance between high resolution and numerical feasibility.
Our simulations account for both the induction fields and the interactions between the magnetospheric plasma and Callisto's atmosphere-ionosphere system.
This combination of major effects on Callisto's magnetic environment has not yet been presented in previous numerical results on Galileo's C03 and C09 flyby.

\begin{figure}
	\noindent\includegraphics[width=\textwidth]{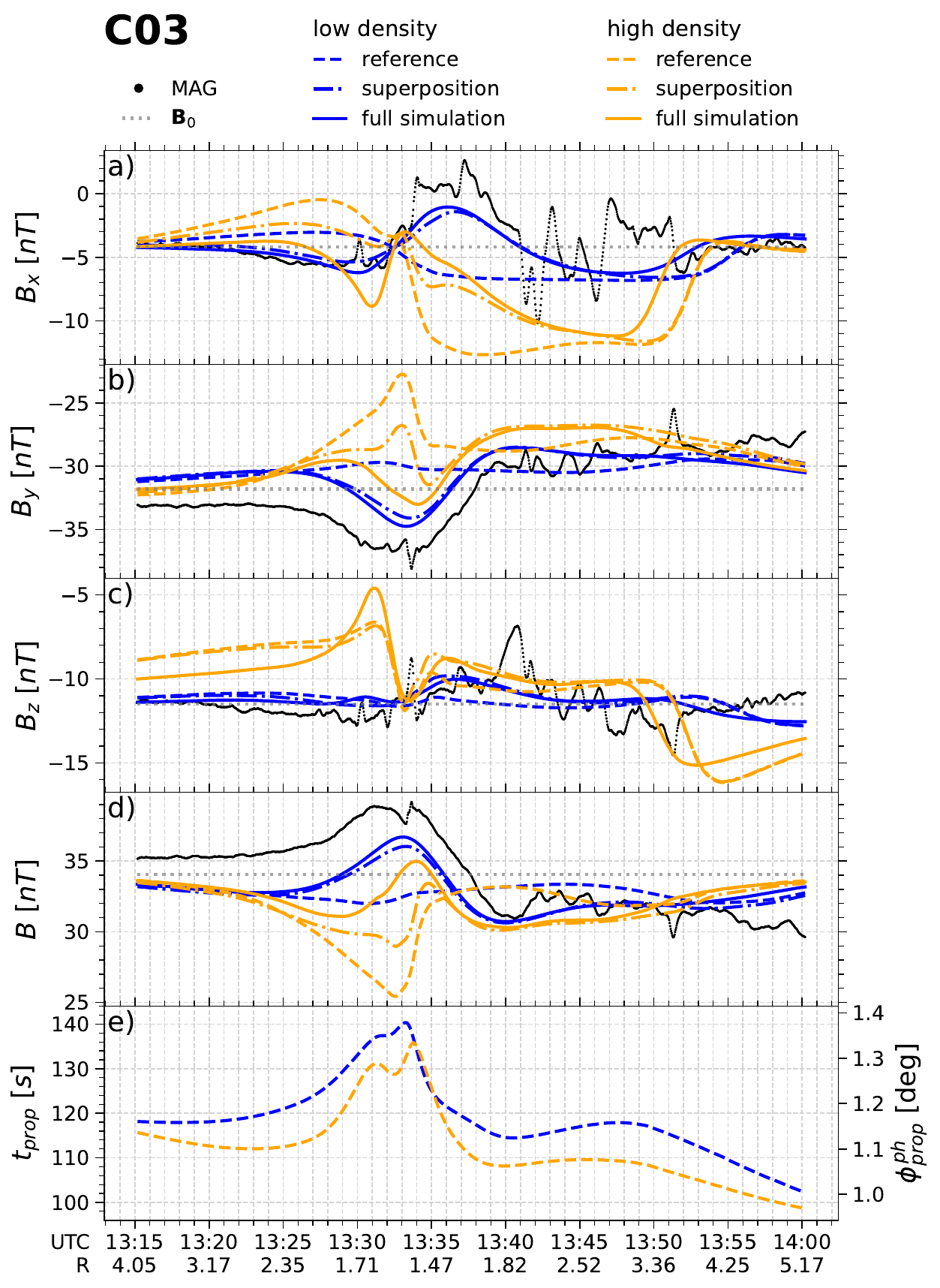}
	\caption{
		Panels a-d: Measured (black) and simulated magnetic field (blue/orange) for Galileo's C03 flyby.
		Panel e: Estimated wave propagation time of the induced magnetic field and the associated phase shift.
	}
	\label{fig:galileo_C03}
\end{figure}

Figure~\ref{fig:galileo_C03} presents the results for the C03 flyby.
As is visible in the reference simulation (dashed orange line) and as is expected from the reacceleration of the plasma \cite<e.g.>[]{Saur2021}, moon-magnetosphere interactions lower the magnitude of the magnetic field in Callisto's wake by about $8\,$nT.
However, this contradicts the Galileo measurements (black dots), which show a ${\sim}4\,$nT increase in magnitude.
On Callisto's downstream side, moon-magnetosphere interactions and induction have an opposite effect on the total magnetic field magnitude. 
Thus, with the superposition of an induced magnetic field (vacuum prediction), the decrease in the magnetic field strength is mostly compensated (dashed-dotted orange line).
While this compensation effect is stronger, when the propagation of the induced magnetic field in the plasma environment and its non-linear feedback is considered (solid orange line), the qualitative fit to the measured data still is poor.

The poor agreement between modelled and measured magnetic field indicates that magnetic perturbations due to moon-magnetosphere interactions were weak during Galileo's C03 flyby.
Thus, we additionally present a model setup with a reduced impact of moon-magnetosphere interactions (dashed blue line).
For this setup, we lowered the upstream plasma mass density by an empirical factor of 10, as this parameter is only poorly constrained.
Note that \citeA{Lindkvist2015} also found their low-density model to better agree with the measurements.
In this low-density model, the peak perturbations due to moon-magnetosphere interactions reduce by ${\sim}6\,$nT with a total magnetic field magnitude that is reduced by only ${\sim}2\,$nT compared to the estimated magnetic background field (gray dotted line). 
With the inclusion of an induced magnetic field, the measurements are well explained qualitatively, although quantitatively our model still underestimates the peak of the magnetic field at about 13:34\,UTC.
This underestimation can be partly attributed to the magnetic background field $\mathbf{B}_0$: while our simulations assume a constant background field, the background field in the Galileo data declines over the course of the ${{\sim}45}$-minute flyby.
In comparison to the vacuum superposition (dashed-dotted blue line), the full simulation result (solid blue line) showcases the stronger signatures of an induced magnetic field.
In the plasma environment, the magnitude of the induced magnetic field appears enhanced by roughly $0.8\,\mathrm{nT} \, \widehat{=}\, 18\,$\%.
Thus, considering the induced magnetic field's propagation with the MHD modes generates results closer to the observations.

We note that the observations obtained during the C03 flyby (black dots) also contain magnetic field perturbations of several nT occuring on spatiotemporal scales of ${\sim}1\,$minute or ${\sim}0.2\,R_\mathrm{C}$, for example in the $B_x$-component at about 13:45\,UTC.
While such fluctuations may already be present in the magnetic field of Jupiter's dynamic magnetosphere, they could also arise from other effects of moon-plasma interactions not included in our MHD model.
These shorter-scale moon-plasma interactions could, for instance, involve electron beams \cite{Mauk2007,Krupp2023} or substructures in the Alfvén wings \cite{Kivelson2004a}.

Based on the geometry of the C03 flyby (\ref{appx:trajectories}), it is likely that Galileo crossed the Jupiter-facing Alfvén wing after its closest approach at 13:34 UTC.
This is consistent with our simulation results, which show a rotation of the direction of the magnetic field in the time interval ${\sim}$13:38-13:51\,UTC, particularly visible as the broad minimum in the $B_x$-component (yellow dashed line).
Because the low-density model has higher Alfvén velocities in the magnetospheric plasma, the Alfvén wing is less inclined, reducing the rotation of the magnetic field (blue dashed line).
In the presence of propagated induced magnetic fields, the system of standing Alfvén waves (the Alfvén wing) is shifted compared to the case without induction \cite{Neubauer1999}. 
For the magnetic field configuration of the C03 flyby, the Jupiter-facing Alfvén wing shifts in the positive $z$-direction, away from the Galileo flyby trajectory.
Accordingly, our full simulations (solid yellow/blue lines) predict that Galileo left the main wing about 2\,minutes earlier compared to the superposition models (dashed-dotted lines), which do not include this effect.

\begin{figure}
	\noindent\includegraphics[width=\textwidth]{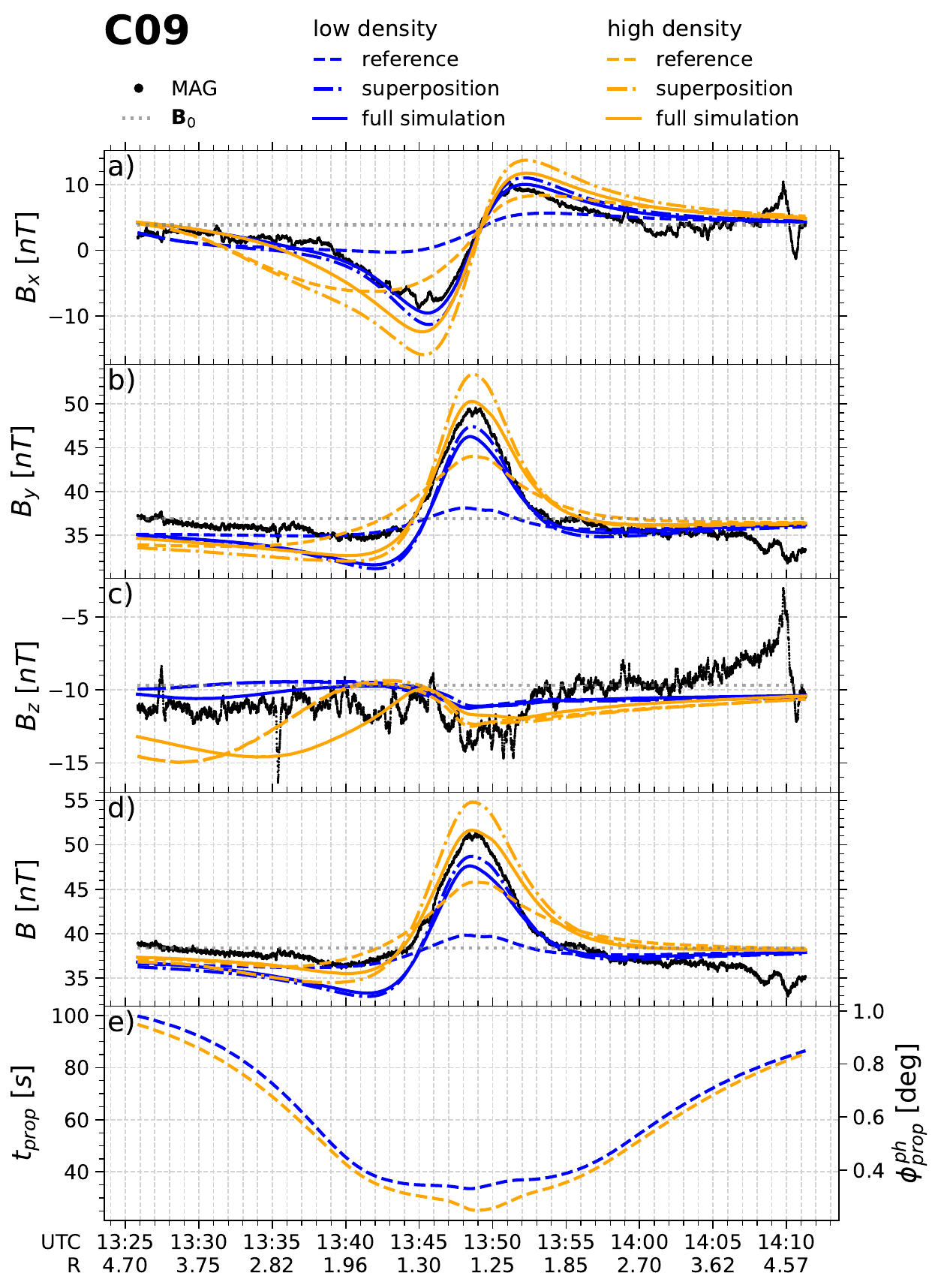}
	\caption{
		Panels a-d: Measured (black) and simulated magnetic field (blue/orange) for Galileo's C09 flyby.
		Panel e: Estimated wave propagation time of the induced magnetic field and the associated phase shift.
	}
	\label{fig:galileo_C09}
\end{figure}

Figure~\ref{fig:galileo_C09} presents the results for the C09 flyby.
On Callisto's upstream side, the magnetic field magnitude is enhanced due to the pile-up effect (solid orange line).
Here, moon-magnetosphere interactions have a somewhat similar imprint on the magnetic field as the induced magnetic field.
Most notably that is the sign reversal in the $B_x$-component and the simultaneous peak in the $B_y$-component close to Galileo's crossing through the plane given by $y=0$, occurring at about 13:48\,UTC.
However, the spatial extent of the magnetic field perturbation due to moon-magnetosphere interactions is broader than supported by the measured data, especially in the $B_x$-component.
The superposition of an induced magnetic field (vacuum prediction) qualitative improves the spatial agreement between measured and modeled magnetic field (dashed-dotted orange line).
However, with an amplitude factor of $A=0.85$, the magnitude of the large scale magnetic field perturbations is overestimated.
Consistent with the results from our simplified setup, our full simulation (solid orange line) predicts reduced perturbations in comparison to the vacuum environment.
Thus, being transported by the MHD modes, the induced magnetic field leaves a smaller imprint on the magnetic field measured during Galileo's C09 flyby compared to transportation by the ordinary electromagnetic wave mode.

To further reduce the magnitude of the overall magnetic field perturbations, two approaches are possible: First, the amplitude of the moon-magnetosphere interactions can be reduced; second, the magnitude of the induced magnetic field can be reduced.
In accordance with the C03 flyby model, we reduce the upstream plasma density by a factor of 10 to reduce the moon-magnetosphere interaction's effect on the magnetic field.
With this approach, the moon-magnetosphere interaction's peak magnetic field perturbations reduce notably by ${\sim}3-5\,$nT (dashed blue line).
Including an induced magnetic field (solid blue line), the low-density model setup can explain the $B_x$-component of the measured magnetic field.
However, while the spatial extent of the simulated magnetic perturbations qualitatively agrees with the measurements, the low-density model underestimates the peak in magnitude at about 13:49\,UTC, mostly determined by $B_y$.

The plasma electron number density $n_e$ is one of the key factors that control the subset of wave modes that can propagate in a given plasma.
It can be expected that both solutions of the induced magnetic field, in a plasma or in a vacuum environment, will be similar for dilute plasmas.
And indeed, both solutions are more similar in the low-density model compared with the high-density model.
With a lower upstream plasma density, the momentum of the plasma reduces, resulting in weaker stresses on the magnetic field.
In turn, moon-magnetosphere interactions perturb Callisto's magnetic field less strongly.
The strength of the magnetic perturbations caused by moon-magnetosphere interactions can be estimated by $\Delta B \approx \overline{\alpha} M_A B_0$ \cite{Saur2013,Saur2021}, with the Alfvén Mach number $M_A$. 
At Callisto, where the ionospheric plasma density strongly dominates over the upstream plasma density, our simulations predict an interaction strength $\overline{\alpha} = |v - v_0|/v_0 > 0.99$, consistent with \citeA{Strobel2002} and \citeA{Liuzzo2015,Liuzzo2016}. 
The estimate therefore reduces to $\Delta B \approx v_0 \sqrt{\mu_0 \rho_0}$.
Consequently, the amplitudes of the magnetic field perturbations are smaller for lower upstream plasma densities and the system behaves more linearly, thereby reducing the transport effects on the induced magnetic field.
Thus, low plasma density cases better justify the description of the induced magnetic field outside the moon by the vacuum superposition.
Nevertheless, even for the low-density model, our results suggest that on Callisto's upstream side, induced magnetic fields have weaker effects as predicted by the vacuum solution.
Vice versa, on Callisto's downstream side, we find that being propagated by the MHD modes, the induced magnetic fields decay less strongly than being propagated by the ordinary electromagnetic mode and thus have a stronger effect than predicted.  

In addition to the quantitative effect of the MHD modes on the induced magnetic field, we also analyze their temporal implications.
The bottom panels of Figures \ref{fig:galileo_C03} and \ref{fig:galileo_C09} respectively show the estimated wave propagation times ((i), see Section~\ref{subsec:results_temporal}) of the induced magnetic field and the associated additional phase shifts for the C03 and C09 flyby.
These times were obtained by integrating the slowness of the fast mode along the straight radial line that is normal to Callisto's surface and connects to the location of measurement (see Equation~(\ref{eq:results_propagation-time})).
For both flybys, the propagation time is not constant. First, because the measurements were obtained at varying distances to Callisto, and second, because the propagation time depends on the non-homogeneously distributed properties of the plasma.
In case of the C09 flyby (Figure~\ref{fig:galileo_C09}), the propagation times and the associated additional phase shifts vary between $25\,$s$ - 100\,$s and $0.2\,^\circ - 1.0\,^\circ$, respectively.
Outside the segment with $R < 2$, a clear dependence on the distance of the measurements is visible.
In case of the C03 flyby (Figure~\ref{fig:galileo_C03}), the propagation times and the associated additional phase shifts are generally higher and vary between $99\,$s$ - 140\,$s and $1.0\,^\circ - 1.4\,^\circ$, respectively.
The largest propagation times occur close to Galileo's closest approach at 13:34\,UTC, in a region of high plasma densities and reduced magnetic field magnitudes.
These results show that, due to transport effects, the location of the measurement requires consideration when interpreting the phase shift of induced magnetic fields.

The times of the arrival of the wave front are exceeded by the times required for the full wave pattern to establish.
For a downstream flyby such as C03, these times also depend on the convection mode.
In the domain $r \leq 1.1\,R_\mathrm{C}$ we compute mean bulk velocities of $v < 0.1\,$km\,\textsuperscript{-1}, resulting in convection times of $R_\mathrm{C}/v > 10^4\,$s.

The C03 and C09 flybys both made their closest approach at a time of declining primary magnetic field magnitudes, i.e., at a time when Callisto was moving towards the magnetic equator of Jupiter. 
Thus, at times previous to the time of the closest approach, the magnitude of the primary magnetic field was larger.
Consequently, with a phase shift ${\phi^\mathrm{ph} > 0}$, the induction signals measured at spacecraft location should have been enhanced compared with the ${\phi^\mathrm{ph} = 0}$ case.
For the C03 flyby and a phase shift of ${\phi^\mathrm{ph}=10^\circ}$, the primary field model of \citeA{Hartkorn2017b} yields a difference of $\Delta B_\mathrm{prim} = 3.1\,$nT in the primary field. 
The associated difference of $\Delta B_\mathrm{sec} \propto A \cdot \Delta B_\mathrm{prim}$, caused by the transport times of the induced magnetic field in Callisto's plasma environment, is another factor to explain the strong induction signals detect by Galileo's magnetometer during the C03 flyby.

\section{Discussion}
\label{sec:discussion}

Our results show that transport effects within the plasma exterior of Callisto have a significant spatiotemporal impact on induced magnetic fields in Callisto's environment.
Here, we briefly discuss how additional sources of asymmetry could affect our results (Section~\ref{subsec:discussion_asymmetry}) and the relevance of transport effects at the other Galilean moons (Section~\ref{subsec:discussion_other-moons}).

\subsection{Additional Sources of Asymmetry}
\label{subsec:discussion_asymmetry}

In Section~\ref{sec:results}, we use a symmetric setup to analyze the propagation effect in isolation and show that transport effects cause a clear upstream/downstream asymmetry in Callisto's induced magnetic field.
In this symmetric setup, the magnetic background field and the dipole moment of the prescribed induced field are perpendicular to the direction of the unperturbed plasma flow.
We lift some of these symmetries and simplifications in Section~\ref{sec:galileo} by introducing a day/night asymmetry in the ionization frequency and by constraining the background and primary magnetic field with the Galileo measurements.
Despite these additional non-symmetric features, the upstream/downstream asymmetry in the induced signals remains and is consistent with the Galileo observations.

Callisto's plasma environment likely possesses additional asymmetries for several reasons, including Callisto's asymmetric atmosphere-ionosphere system \cite{Hartkorn2017a,Vorburger2019,Mogan2021} and non-symmetric moon-magnetosphere interactions associated with non-MHD effects \cite{Liuzzo2015,Liuzzo2016,Liuzzo2017}.
However, as long as the total magnetic field is approximately perpendicular to the direction of the corotational plasma flow, the upstream/downstream asymmetry in the induced fields should persist.
This is because under these conditions, the transport of the induced signals on the upstream side remains limited to the fast mode, while on the downstream side, the induced signals additionally propagate with the convection mode.
In Callisto's almost linearly polarized primary magnetic field, the $B_y$ component dominates the $B_x$ component \cite{Seufert2011,Vance2021}.
Judging from the data in Figure 1b of \citeA{Kivelson1999}, the background magnetic field is never rotated more than $16^\circ$ in the $\pm x$-direction.
Therefore, we expect the upstream/downstream asymmetry to be a persistent feature of the induced magnetic fields at Callisto.
Since the propagation of the MHD modes depends on the properties of the plasma environment, the additional asymmetries in Callisto's plasma environment will add another layer of complexity to the propagation of the induced magnetic field. 
Consequently, for a quantitative interpretation and in-depth understanding of the induction signals, the transport effects need to be constrained individually for the respective plasma configuration of a flyby.

\subsection{Transport Effects at the Other Galilean Moons}
\label{subsec:discussion_other-moons}

The induced magnetic fields of Io, Europa, and Ganymede are also subject to transport effects in the external space plasma environment. As these moons are exposed to a denser magnetospheric plasma and a stronger magnetic field, the conditions for the propagation of the induced magnetic fields with the MHD modes are clearly fulfilled.
Among the Galilean satellites, Io and Callisto, the two moons with the densest atmospheres, feature the highest interaction strengths $\overline{\alpha}$ \cite{Saur2013}.
This is also reflected in the values listed in \citeA{Bagenal2020}, based on which relative magnetic field perturbations caused by moon-magnetosphere interactions are on the order of $37\,\%$ and $18\,\%$ at Io and Europa, respectively.
Thus, because of the higher degree of non-linearity, we would expect transport effects at Io similar to Callisto, but smaller at Europa.
However, robust estimates of the importance of the transport effects require numerical models and are beyond the scope of our Callisto-focused study.
This is especially true for Ganymede, where the moon's dynamo magnetic field further complicates the interaction scenario.

\section{Summary and Conclusion}
\label{sec:conclusion}

Due to the low frequency of the inducing fields, we showed that the induced magnetic field cannot propagate with the ordinary electromagnetic mode in Callisto's plasma environment.
Instead, the induced magnetic field propagates with the MHD wave modes \textit{slow}, \textit{fast}, and \textit{Alfvén} and the \textit{convection} mode, which has important consequences when interpreting magnetic field measurements.
In this work, we investigated the effects of this difference in propagation.

Using an MHD model, we simulated the coupled system of induced magnetic fields and plasma interactions of Jupiter's magnetospheric plasma with the atmosphere/ionosphere of Callisto.
We compared these model results with the induced magnetic field that is expected if it would propagate with the ordinary electromagnetic wave mode.
Our results show that the modes of propagation have spatial as well as temporal implications on the observable induced magnetic field.

Being propagated by the anisotropic MHD modes, the induced magnetic fields in Callisto's plasma environment include asymmetric properties with a pronounced upstream/ downstream asymmetry.
In Callisto's upstream hemisphere, the induced magnetic field is weaker than expected from considerations of a vacuum environment.
In Callisto's downstream hemisphere, in contrast, the induced magnetic field decays less strongly than expected.
Therefore, on Callisto's downstream side, the induced magnetic field is observable at larger distances.
We find that by neglecting the effects of the propagation, the amplitude of the induced field can be misjudged:
On Callisto's upstream side, the strength of the induced magnetic field is underestimated, while on the downstream side, it is overestimated.
This effect of over- or underestimation is largely independent of the amplitude of the induced magnetic field, but it depends on the strength of the moon-magnetosphere interactions. 
If magnetic perturbations caused by moon-magnetosphere interactions are low, the transport effects weaken and vice versa.
In our model study, which aims to reflect the plasma conditions for positions of Callisto above or below the Jovian plasma sheet, the effect of under- or overestimating amounts to about 26\,\% and 39\,\%, respectively.

Within Callisto's plasma environment, MHD wave and convection velocities are strongly reduced compared to their upstream values.
The reduced propagation velocities introduce a temporal delay between the emergence of the induced magnetic field at the surface of the moon and its arrival at spacecraft location.
The associated phase shift is in addition to the phase shift from the induction process and possesses a spatial dependency because the MHD wave velocities depend on the properties of the local plasma and these properties are not homogeneously distributed.
Transport phase shifts can exceed several degrees and are larger for downstream than for upstream flybys.
In regions that are accessible not only to the fast mode but also to the convection mode, maximum transport time scales can reach orders of ${>10^3}\,$s (${>10^\circ}$).

Our simulation results for the C03 and C09 flybys show that the transport effects are consistent with Galileo's magnetic field measurements and suggest that upstream plasma mass densities should have been low at the time of the flybys.
Including transport effects improves the qualitative agreement with the observations compared with our superposition model, which neglects these effects.
This is especially true for the C03 downstream flyby, where moon-magnetosphere interactions reduce the magnetic field strength, thus requiring relatively large amplitudes of the induced magnetic field.
At the upstream C09 upstream flyby, on the other hand, moon-magnetosphere interactions contribute with a somewhat similar imprint to the observed magnetic field perturbations as the induced magnetic field.
In combination of these effects, the amplitude $A$ of Callisto's induced magnetic field may not need to be as large as suggested by analytical estimates to explain the observations, thus supporting the possibility of a deeper buried ocean or the ionosphere as the source region of Callisto's induction response.

The observation and interpretation of induced magnetic fields is a valuable tool to constrain the electrically conductive parts of a planetary body such as the ionosphere or a subsurface ocean. 
Exhibiting anisotropic propagation properties and finite propagation velocities, the MHD modes have a non-negligible spatiotemporal impact on the induced magnetic field.
Therefore, analyzing the propagation of the induced magnetic field at Callisto and the other Galilean moons will be important for the upcoming Europa Clipper and JUICE missions for an advanced understanding of induction effects. 


\appendix

\section{Model Parameters and Simulation Setup}
\label{appx:values}

To conduct our simulations, we require a set of values that characterizes the properties of the inflowing magnetospheric plasma and of Callisto's neutral gas envelope.
We motivate our particular choice of values in Sections \ref{subappx:values_plasma} and \ref{subappx:values_callisto}.
A description of our simulation setup and procedure is included in Section~\ref{subappx:values_numerics}.

\subsection{Model Parameters: Upstream Plasma Conditions}
\label{subappx:values_plasma}

In our simulations, we assume the upstream magnetospheric plasma to be spatially homogeneous and constant in time for each run.
Thus, one value each for $\rho_0$, $\mathbf{v}_0$, $\mathbf{B}_0$ and $p_0$ is required.

The plasma mass density is the product of the plasma number density and the average ion mass: $\rho_0 = n_0 \cdot m_i$.
We choose an ion mass of $m_i = 16$\,amu.
This is consistent with the mean (equatorial) value listed by \citeA{Kivelson2004} and with an  O\textsuperscript{+} dominated plasma.
\citeA{Kim2020} identified H\textsuperscript{+} and S\textsuperscript{++} as the most abundant ion species for the off-equator plasma in the region between $25 - 30$ Jupiter radii ($R_J$).
However, their mean ion mass of $\overline{m}_i = 18.6$\,amu does not significantly differ from our value.
Plasma number densities at Callisto are poorly constrained \cite{Kleer2023}.
Based on the functional model of \citeA{Bagenal2011} and a current sheet distance of $z = 3.5 R_J$, we set $n_0 = 0.06\,$cm\textsuperscript{-3}. 
This results in $\rho_0 = 0.96$\,amu\,cm\textsuperscript{-3}.

At Callisto's orbit, the magnetospheric plasma is no longer in rigid corotation \cite{Bagenal2016}. 
We assume the plasma to flow in azimuthal direction only and adapt a bulk velocity of $\mathbf{v}_0 = 192 \hat{\mathbf{e}}_x$\,km\,s\textsuperscript{-1} \cite{Kivelson2004}.

For the thermal pressure, which is poorly constrained at Callisto, we adopt a value of $p_0 = 9.6 \cdot 10^{-2}\,$nPa. 
Together with magnetic field magnitudes of $B_0 < 40$\,nT, this value gives a plasma $\beta = p/(B^2/2\mu_0) \geq 0.15$, which is on the same order of magnitude as listed by \citeA{Kivelson2004} and is the main reason for this selection.

At positions, when Callisto is located far above or below the plasma sheet, the radial component of Jupiter's magnetospheric magnetic field dominates.
To simplify the geometry of the interaction scenario, we neglect the latitudinal and azimuthal component and select $\mathbf{B}_0 = 35\hat{\mathbf{e}}_y$\,nT in our simplified scenario.
For the C03 and C09 flyby scenario, we derive the values for the upstream magnetic field directly from the Galileo magnetometer measurements \cite{Kivelson1997a}.
For this, we fit second order polynomials to the magnetic field components measured in distances between $5-10\,R_\mathrm{C}$ before and after the closest approach. 
Evaluating the polynomials at the position of the closest approach yields $\mathbf{B}_0 = \left(-4.2, -31.8, -11.5\right)$\,nT and $\mathbf{B}_0 = \left(3.9, 36.9, -9.7\right)$\,nT for the C03 and C09 flyby, respectively.
As at Callisto, the inducing or primary magnetic field mainly is confined to its $xy$-plane \cite{Seufert2011}, we set $\mathbf{B}_\mathrm{prim} = B_{0,x}\hat{\mathbf{e}}_x + B_{0,y}\hat{\mathbf{e}}_y$. 

\subsection{Model Parameters: Callisto}
\label{subappx:values_callisto}

Callisto's atmosphere likely is O\textsubscript{2} dominated \cite{Cunningham2015}.
Column densities derived from telescope or radio occultation observations range between $~4 \cdot 10^{15}$\,cm\textsuperscript{-2} \cite{Cunningham2015,Kleer2023} and $~4 \cdot 10^{16}$\,cm\textsuperscript{-2} \cite{Kliore2002}.
\citeA{Hartkorn2017a} modelled Callisto's atmosphere constrained by both, the ultraviolet brightnesses observed by \citeA{Cunningham2015} and the electron densities observed by \citeA{Kliore2002}. 
In our model, we adopt their mean column density of $N_n \approx 2.1 \cdot 10^{15}$\,cm\textsuperscript{-2} and use $m_n = 32$\,amu. 
Realistic scale heights would probably be on the order of $~30$\,km \cite{Hartkorn2017a}.
However, grid resolutions capable to resolve such a scale height would heavily increase the numerical workload. 
Thus, in analogy with previous models of Callisto's moon-magnetosphere interactions \cite{Seufert2012,Liuzzo2015}, we use an artificial enhanced scale height of $H = 230$\,km for our simplified scenario.
For the flyby scenarios we improve on the resolution and set $H = 60\,$km.

In our model, the plasma population gains energy from the atmosphere through thermally driven ion-neutral collisions and by ionization.
These processes are mildly influenced by the temperature $T_N = 300\,$K, which is higher than the temperature of $150 \pm 50\,$K reported by \citeA{Carlson1999} for the CO2 component of Callisto's atmosphere.
However, we expect plasma and Joule heating to enhance Callisto's atmospheric temperature at higher altitudes similar to expectations at Io \cite{Strobel1994}.
As tested with our symmetric reference model, the choice of $T_N$ only has a small impact on the simulation results ($\overline{\Delta B} < 3\cdot 10^{-4}\,$nT in the domain $r \leq 3\,R_\mathrm{C}$, compared with $T_N=150\,$K).

At Callisto, photoionization is an important ionization mechanism \cite{Kliore2002,Cunningham2015,Hartkorn2017a}.
Electron impact ionization is an additional, but poorly constrained ionization mechanism \cite{Kleer2023,Mogan2023}.
For an O\textsubscript{2} atmosphere and a medium active Sun, the photoionization frequency is $\nu_\mathrm{ion, ph} \approx 3.0 \cdot 10^{-8}$s\textsuperscript{-1}
(obtained from the online database of \citeA{Huebner2015} and scaled to Jupiter's average Sun distance of $5.2$\,AU).
During both, the C03 and C09 flyby, the sub-solar point was on Callisto's leading (downstream) hemisphere at $\phi = 250.0^\circ$ and $\phi=277.0^\circ$, respectively \cite{Seufert2012}.
For the flyby scenarios (Section~\ref{sec:galileo}), we use $\nu_\mathrm{ion} = \nu_\mathrm{ion, ph}$ at the sunlight hemisphere.
On Callisto's night side, which roughly coincided with its trailing (upstream) hemisphere, we retain a value of $\nu_\mathrm{ion} = 0.1 \cdot \nu_\mathrm{ion, ph}$ to account for the possible electron impact ionization.
In the case of the simplified scenario (Section~\ref{sec:results}), we omit any day/night asymmetries in the ionization frequency.
Since we are interested in the asymmetries due to the anisotropic wave propagation, we minimize additional asymmetries in this part of the study.
Further, we adopt an artificially low value of $\nu_\mathrm{ion} = 1/25 \cdot \nu_\mathrm{ion, ph}$ in the simplified scenario.
With a regular ionization rate a dense ionosphere builds up. 
In our model we set the induced magnetic field at the surface of Callisto and a dense ionosphere would effectively suppress the propagation of the induced magnetic field because the MHD wave velocities are strongly dependent on the plasma density.
In such a case, however, a significant portion of the induction process may occur in the dense ionosphere directly \cite{Hartkorn2017b}.
Rather than from the surface of Callisto, in this case, the induced magnetic field would propagate from the top of the ionosphere.
For this reason, the artificially low ionization frequency is an attempt to cover both extreme cases: when the ionosphere is thin and the possible subsurface ocean is the source region of the induced field, and when the ionosphere is dense and the source region of the induced field. 

\subsection{Numerical Solver and Setup}
\label{subappx:values_numerics}

We carry out our numerical simulations with PLUTO, a finite volume code specialized in modeling astrophysical fluids \cite{Mignone2007}.
For our study we select PLUTO's piece-wise linear, 2nd order accurate scheme for the reconstruction of the variables at the cell interfaces, the Harten, Lax, van Leer (HLL) approximate solver for the Riemann problem and a 2nd order Runge-Kutta method for the integration forward in time.

Using spherical geometry, we split the 3D-simulation domain into $384 \times 112 \times 192 \left(r,\theta,\phi \right) \approx 8 \cdot 10^6$ cells. 
Making use of a non-equidistant grid in radial direction, we archive a resolution of $0.02\,R_\mathrm{C}$ below $r=1.3\,R_\mathrm{C}$, of less than $0.066\,R_\mathrm{C}$ below $r=5\,R_\mathrm{C}$ and of a minimum of $1.35\,R_\mathrm{C}$ at the outer boundary at $r=100\,R_\mathrm{C}$.
To avoid overly strict time step restrictions at the pole axis singularity, starting below (above) $\theta = 45^\circ (135^\circ)$, we gradually decrease the latitudinal resolution from its minimum value of $1.5^\circ$ to a value of $1.96^\circ$.
Additionally, we make use of PLUTO's ring average method, in which respectively 4 and 2 cells in the first two rings around the pole are grouped in the calculation.
In longitudinal direction, we employ a constant resolution of $1.875^\circ$.
To resolve smaller scale heights in the flyby models, we improve the grid size to $400 \times 120 \times 240 \left(r,\theta,\phi \right)$ cells, yielding a minimal radial resolution of $0.008\,R_\mathrm{C}$.

We divide the outer boundary into two sections: on the upstream side ($\phi \leq 180^\circ$), we use fixed boundary values characteristic for the inflowing plasma (see Table~\ref{tab:model_values} or Section~\ref{subappx:values_plasma}); and on the downstream side ($\phi > 180^\circ$), we use open boundary conditions. 
At the inner boundary at $r = 1\,R_\mathrm{C}$, we employ the boundary conditions derived by \citeA{Duling2014}.
In these conditions, Callisto is treated as having an electrically non-conducting, plasma-absorbing surface, which is realized by open boundary conditions, the criterion $v_r \leq 0$ and a special treatment of the magnetic field.
The treatment of the magnetic field allows the inclusion of an induced magnetic field whose values are set at the inner boundary.


\section{Galileo Flyby Trajectories}
\label{appx:trajectories}

\begin{figure}
	\noindent\includegraphics[width=\textwidth]{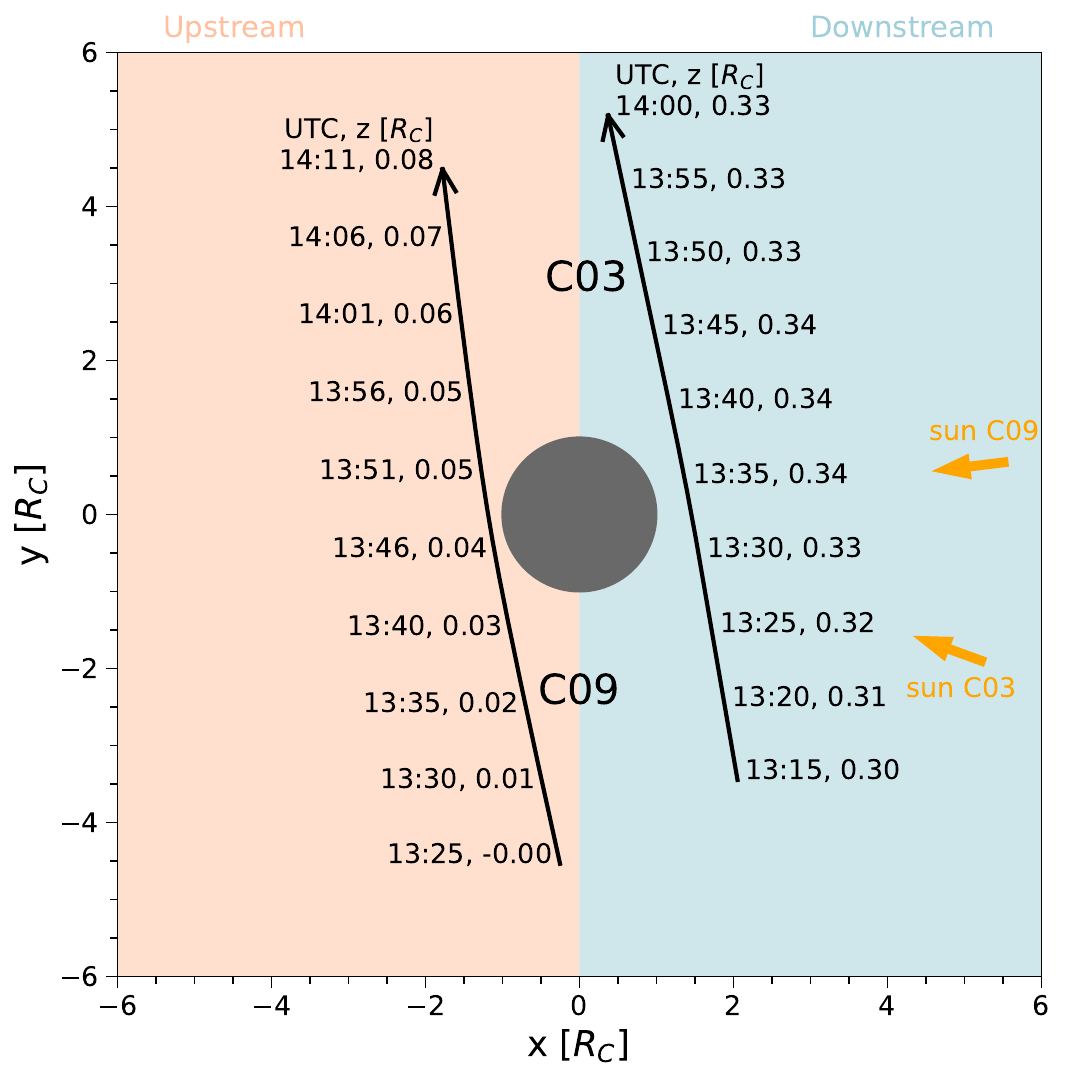}
	\caption{
        Flyby trajectories of Galileo's C03 and C09 flyby projected onto the $xy$-plane.
        The time and the $z$-coordinates are provided as ticks next to the trajectories.
        The orange arrows indicate the direction of the Sun for the respective flyby.
	}
	\label{fig:trajectories_trajectories}
\end{figure}

Figure~\ref{fig:trajectories_trajectories} visualizes the flyby trajectories of Galileo's C03 and C09 flyby at Callisto.
During both flybys, Galileo approached Callisto from the Jupiter-averted side and traveled in a predominantly positive y-direction, with a slight inclination in the upstream direction. 
During C03, Galileo remained in the downstream hemisphere at an average distance of $z=0.33\,R_\mathrm{C}$ above Callisto's equatorial plane. In contrast, during C09, Galileo passed through the upstream hemisphere at a distance $z < 0.1\,R_\mathrm{C}$ from the equatorial plane.
In both flybys, the Sun mainly illuminated Callisto's downstream hemisphere.


\section*{Open Research Section}

The PLUTO code used for this work is freely available at https://plutocode.ph.unito.it/ (version 4.4).
The Galileo magnetometer observations for the C03 and C09  Callisto flyby are publicly available on NASA's Planetary Data System at https://doi.org/10.17189/1519667 \cite{Kivelson1997a} with product IDs ORB03\_CALL\_SYS3 and ORB09\_CALL\_SYS3.
The simulation output data of our symmetric model as well as of the C03 and C09 flyby models are available in a Zenodo repository at https://doi.org/10.5281/zenodo.13152396 \cite{Strack2024} with CCA 4.0 license.

\acknowledgments
This project has received funding from the European Research Council (ERC) under the European Union's
Horizon 2020 research and innovation program (grant agreement no. 884711).
This work used resources of the Deutsches Klimarechenzentrum (DKRZ) granted by its Scientific Steering Committee (WLA) under project ID 1350.
We furthermore thank the Regional Computing Center of the University of Cologne (RRZK) for computing time on the DFG-funded (Funding number: INST 216/512/1FUGG) High Performance Computing system CHEOPS as well as support.


\end{document}